\documentclass[10pt,final, twocolumn]{IEEEtran}
\IEEEoverridecommandlockouts

\usepackage{cite}
\usepackage{amsmath,amssymb,amsfonts}
\usepackage{algorithmic}
\usepackage{graphicx}
\usepackage{textcomp}
\usepackage{xcolor}
\def\BibTeX{{\rm B\kern-.05em{\sc i\kern-.025em b}\kern-.08em
    T\kern-.1667em\lower.7ex\hbox{E}\kern-.125emX}}

\usepackage{acronym}
\usepackage[belowskip=-10pt,aboveskip=1.8pt]{caption}
\usepackage{standalone}
\usepackage{subcaption}
\usepackage{tikz}
\usepackage{url,enumitem, cite}
\usepackage{verbatim}
\usepackage[bookmarks,colorlinks]{hyperref}
\usepackage{soul, xcolor}
\usepackage{mathrsfs}
\usepackage[ruled,linesnumbered,vlined]{algorithm2e}
\SetKwInput{KwData}{\textbf{Init}} 
\usepackage{adjustbox}
\usepackage{multirow}
\usepackage{makecell}
\usepackage{multirow,colortbl,booktabs} 
\usepackage{xcolor-material}
\usepackage{lscape}

\usepackage[all=normal,paragraphs=tight,floats=normal,mathspacing=normal,wordspacing=tight,charwidths=tight,mathdisplays=normal,leading=normal]{savetrees}

\newcommand{\myVec}[1]{{\boldsymbol{#1}}}
\newcommand{\myMat}[1]{{\boldsymbol{#1}}}
\newcommand{\mySet}[1]{\mathcal{#1}}

\newcommand{\thresh}{\Lambda}
\newcommand{\name}{NF-SubspaceNet}

\let\oldnl\nl
\newcommand{\nonl}{\renewcommand{\nl}{\let\nl\oldnl}}

\acrodef{ai}[AI]{artificial intelligence}
\acrodef{dl}[DL]{deep learning}
\acrodef{bs}[BS]{base station}
\acrodef{dnn}[DNN]{deep neural network}
\acrodef{cnn}[CNN]{convolutional neural network}
\acrodef{dcnn}[DCNN]{deconvolutional neural network}
\acrodef{mlp}[MLP]{multi-layer perceptron}
\acrodef{snr}[SNR]{signal-to-noise ratio}
\acrodef{awgn}[AWGN]{additive white Gaussian noise} 
\acrodef{ml}[ML]{machine learning} 
\acrodef{sgd}[SGD]{stochastic gradient descent} 
\acrodef{mse}[MSE]{mean-squared error}
\acrodef{rmse}[RMSE]{root mean squared error}
\acrodef{rmspe}[RMSPE]{root mean squared periodic error}
\acrodef{mle}[MLE]{maximum likelihood estimation}
\acrodef{snr}[SNR]{signal-to-noise ratio}
\acrodef{admm}[ADMM]{alternating direction method of multipliers}
\acrodef{aoa}[AoA]{Angle of Arrival}
\acrodefplural{aoa}[AoAs]{Angles of Arrival}
\acrodef{em}[EM]{electromagnetic}
\acrodef{cmos}[CMOS]{complementary metal-oxide semiconductor}
\acrodef{ula}[ULA]{uniform linear array}
\acrodef{em}[EM]{Electromagnetic}
\acrodef{doa}[DoA]{direction of arrival}
\acrodef{music}[MUSIC]{MUltiple SIgnal Classification}
\acrodef{esprit}[ESPRIT]{Estimation of signal parameters via rotational invariance techniques}
\acrodef{rmusic}[R-MUSIC]{Root-MUSIC}
\acrodef{evd}[EVD]{eigenvalues decomposition}
\acrodef{sps}[SPS]{spatial smoothing}
\acrodef{iid} [i.i.d] {Independent and identically distributed}
\acrodef{ls}[LS]{least squares}
\acrodef{relu}[ReLu]{Rectified Linear Unit}
\acrodef{crb} [CRB] {Cramér–Rao Bound}
\acrodef{ccrb} [CCRB] {Conditional Cramér–Rao Bound}
\acrodef{dcdmusic}[DCD-MUSIC]{deep-learning-aided cascaded differentiable MUSIC}
\acrodef{transmusic}[TransMUSIC]{Transformer MUSIC}
\acrodef{pmn}[PMN]{PeakMUSICNet}
\acrodef{ssn}[SSN]{SubSpaceNet}
\acrodef{mdl}[MDL]{Minimum Description Length}
\acrodef{aic}[AIC]{Akaike Information Criterion}
\acrodef{ae}[AE]{auto encoder}
\acrodef{mac}[MACs]{Multiply-Accumulate Operations}

\setstcolor{blue}
\definecolor{mygreen}{RGB}{0, 150, 0}
\newcommand{\bgreen}[1]{\textcolor{mygreen}{\textbf{#1}}}

\begin{document}

\title{Near Field Localization via AI-Aided Subspace Methods}

\author{Arad Gast, Luc Le Magoarou, and Nir Shlezinger
\thanks{Parts of this work were presented at the IEEE International Conference on Acoustics, Speech, and Signal Processing (ICASSP) 2025 as the paper \cite{gast2025dcdmusic}.
A. Gast and N. Shlezinger are with the School of ECE, Ben-Gurion University of the Negev, Be’er-Sheva, Israel (e-mails: gast@post.bgu.ac.il;
nirshl@bgu.ac.il). L. Le Magoarou is with Univ. Rennes, INSA Rennes, CNRS, IETR-UMR 6164, France (e-mail: luc.Le-magoarou@insa-rennes.fr). The work was supported  by the European Research Council (ERC) under the ERC starting grant nr. 101163973 (FLAIR), the Israel Innovation Authority, and  the French national research agency, grant ANR-23-CE25-0013.}

}



\maketitle
\begin{abstract}
The increasing demands for high-throughput and energy-efficient wireless communications are driving the adoption of extremely large  antennas operating at high-frequency bands. In these regimes, multiple users will reside in the radiative near-field, and accurate localization  becomes essential. Unlike conventional far-field systems that rely solely on \ac{doa} estimation, near-field localization exploits spherical wavefront propagation to recover both \ac{doa} and range information. While subspace-based methods, such as \ac{music} and its extensions, offer high resolution and interpretability for near-field localization, their performance is significantly impacted by model assumptions, including non-coherent sources, well-calibrated arrays, and a sufficient number of snapshots. To address these limitations, this work proposes \ac{ai}-aided subspace methods for near-field localization that enhance robustness to real-world challenges. Specifically, we introduce {\em NF-SubspaceNet}, a deep learning-augmented 2D MUSIC algorithm that learns a surrogate covariance matrix to improve localization under challenging conditions, and {\em DCD-MUSIC}, a cascaded \ac{ai}-aided approach that decouples angle and range estimation to reduce computational complexity. We further develop a novel model-order-aware training method to accurately estimate the number of sources, that is combined with casting of near field subspace methods as \ac{ai} models for learning. Extensive simulations demonstrate that the proposed methods outperform classical and existing deep-learning-based localization techniques, providing robust near-field localization even under coherent sources, miscalibrations, and few snapshots.
\end{abstract}
%

\acresetall
 
\section{Introduction}
Wireless communications systems are subject to constantly growing demands in terms of throughput, coverage, and energy efficiency. Several emerging technologies are expected to be combined in order to meet these demands~\cite{saad2019vision}. One of the  technologies that are envisioned to play a key role in future wireless systems involves employing extremely large~\cite{wang2024tutorial} and holographic arrays~\cite{huang2020holographic} at the \ac{bs}, using, e.g., large intelligent surfaces~\cite{hu2018beyond} and dynamic metasurface antennas~\cite{shlezinger2021dynamic}. Concurrently, high-frequency bands, such as millimeter waves~\cite{rappaport2019wireless} and sub-THz~\cite{jiang2024terahertz} regimes are expected to be explored in order to utilize their abundant bandwidth. 

The combination of extremely large surface based antennas with high frequency signaling  gives rise to two considerations: 
$(i)$ accurate localization of mobile devices becomes essential, allowing  \acp{bs} to generate the focused beams that are enabled by large surfaces and are crucial at high frequencies~\cite{jiang2024terahertz};  
$(ii)$ the Fraunhofer limit  can be in the order of over tens of meters~\cite{zhang2022beam}, and thus some of the users  will likely reside in the {\em radiative near-field} (termed {\em near-field} henceforth)~\cite{zhang20236g, lu2024tutorial}. 

The fact that some of the users are located in the near-field gives rise to passive localization capabilities that are not present in the conventional far-field. While far-field narrowband sources can only be localized in terms of their \ac{doa},    near-field spherical wavefronts allow recovering both \ac{doa} and range~\cite{elzanaty2023toward}. Algorithmic tools for localizing multiple near-field sources  employ beamforming~\cite{yang2023near}, maximum-likelihood computation~\cite{cheng2022efficient}, compressive sensing techniques~\cite{hu2014near, rinchi2022compressive}, cumulant-based schemes~\cite{Guanghui2020,Challa1995HighOrderEsprit}, null-space projection~\cite{sun2024projection}, and covariance-based subspace methods~\cite{2dmusic1991,zuo2020subspace,  zhang2018localization, Liang2010MixFarNearLocalization,zhi20071dmusicesprit,Starer1994pathfollow,ebadi2024near}. The latter family encompasses extensions of the \ac{music} algorithm~\cite{schmidt1986music} to near-field signals via, e.g., 2D \ac{music}~\cite{2dmusic1991,zuo2020subspace}, multi-stage \ac{music}~\cite{zhang2018localization, Liang2010MixFarNearLocalization,zhi20071dmusicesprit,Starer1994pathfollow}, and iterative 1D \ac{music}-type methods~\cite{ebadi2024near}. The family of subspace methods is  attractive as: $(i)$ the resolution of such algorithms is not dictated by the array geometry; $(ii)$ they naturally support recovering multiple sources;  $(iii)$ they are interpretable, in the sense of providing a meaningful spectrum representation; and $(iv)$ they  are often simple to implement. However, near-field subspace methods rely on several modeling assumption, e.g., non-coherent sources, calibrated arrays, and many snapshots; when these are violated, performance degrades considerably.

An alternative  approach, which does not rely on modeling assumptions,   learns to localize from data. Such methods rely on the recent advances in \ac{ai} tools, and particularly deep learning~\cite{chen2020survey}. Various \acp{dnn} were suggested for such tasks, including \acp{mlp}~\cite{DNN_WITH_Antenna_ARRAY, cong2020robust}, \acp{cnn}~\cite{cao2020complex, DOAEstimation_LowSNR, DOAEstimation_SparsePrior, lee2022deep, qin2023deep,jiang2022deep, weisser2023unsupervised,su2021mixed}, and attention models~\cite{lan2023novel,ji2024transmusic}. While black-box architectures rely on  non-interpretable highly parameterized architectures trained with massive data sets, \acp{dnn} can also be combined with classic localization algorithms via model-based deep learning~\cite{shlezinger2023model}. For far-field \ac{doa} recovery, several different augmentations of   subspace methods with \acp{dnn} were proposed~\cite{lee2022ftmr,elbir2020deepmusic, barthelme2021doa, wu2022gridless, jiang2023toeplitz, DA-MUSIC-2023,xu2024md, shmuel2024subspacenet}. These include training \acp{dnn} to produce \ac{music} spectrum~\cite{lee2022ftmr,elbir2020deepmusic},  and the usage of \ac{dnn} models to compute a covariance matrix, either to approach some ground truth clear covariance as in \cite{barthelme2021doa, wu2022gridless, jiang2023toeplitz}, or to constitute a surrogate covariance for downstream \ac{doa} recovery~\cite{DA-MUSIC-2023,xu2024md, shmuel2024subspacenet}.  The latter approach enables applying  subspace methods without being confined to, e.g., coherent sources. However, these existing \ac{ai}-aided methods  are  geared towards far-field sources, and rely on differentiable \ac{doa} estimation, which does not extend to the near-field. 

This work proposes \ac{ai}-aided  methods that learn to perform subspace-based near-field localization. We aim to preserve the interpretable operation of subspace near-field localizers, and particularly the ability to obtain an informative eigen and \ac{music} spectra that enable identifying the number of sources and their location, while coping with coherent sources, miscalibrations, and few snapshots.

Our main contributions are summarized as follows:
\begin{itemize}
  \item  {{\bf AI-Aided Near-Field \ac{music}:} We propose a deep learning-aided extension of 2D near-field \ac{music}, termed {\em \name}. Our method follows the general methodology of \acl{ssn}~\cite{shmuel2024subspacenet}, which learns to produce a surrogate covariance matrix tailored for subspace-based localization. However, \acl{ssn} relies on the differentiability of far-field estimators such as \acl{rmusic} during training, which are not applicable in the near-field setting. To enable end-to-end training in this more challenging regime, we introduce a differentiable loss function inspired by~\cite{chatelier2024physically}, which supervises the \ac{music} spectrum itself rather than the final estimated parameters, thus avoiding the need for differentiable post-processing.
}
    \item {\bf Reduced Complexity Cascaded Algorithm:} As 2D \ac{music}, either model-based or \ac{ai}-aided, involves a two-dimensional grid search that can be computationally intensive, we propose an alterative reduced complexity  algorithm termed {\em \ac{dcdmusic}}. \ac{dcdmusic} is based on \ac{esprit}-\ac{music} cascade~\cite{zhi20071dmusicesprit}, decomposing \ac{doa} and range recovery, leveraging the differentiability of \acs{esprit} to train a dedicated \ac{dnn} to compute a surrogate far-field covariance, used to recover the \acp{doa}. We then propose a soft-to-hard approximation of \ac{music}  to train another \ac{dnn} to produce a covariance  for range recovery via 1D \ac{music} for each \ac{doa}. %
    \item {\bf Model-Order-Aware Training Method:} We introduce a novel training scheme that combines the accuracy in localizing the sources with the need to {\em identify the number of sources}. We specifically propose three different training loss terms that balance the ability to recover the number of sources from the surrogate eigenspectrum via conventional thresholding, as well as model-order selection techniques based on \ac{mdl}~\cite{rissanen1978modeling} and \ac{aic}~\cite{akaike1998information}.  
     \item {\bf Extensive Experimentation}: We provide a detailed qualitative and quantitative evaluation of our algorithms. For the latter, we contrast  our \ac{ai}-aided methods with representative model-based and data-driven near-field localization methods. We demonstrate that our methods outperform various benchmarks and  localize multiple  near-field as well as far-field coherent sources in challenging settings while providing a meaningful and interpretable spectrum. 
\end{itemize}

The rest of this paper is organized as follows: Section~\ref{sec:System Model and Preliminaries} formulates  the near field localization setup, and briefly reviews relevant subspace methods. Section~\ref{sec:method} presents our \ac{ai}-aided near-field localization methods, 
 that are  numerically evaluated in Section~\ref{sec:experiments}, while Section~\ref{sec:conclusions} provides concluding remarks.

Throughout this paper, we use boldface lower-case and upper-case letters for vectors (e.g., $\myVec{x}$) and matrices (e.g., $\myMat{X}$), respectively. The $n$th entry of  $\myVec{x}$ is denoted by $[\myVec{x}]_n$. Calligraphic letters denote sets, e.g., $\mySet{X}$, with $\mathbb{C}$ being the set of complex numbers,  while $\|\cdot\|$,  $\myVec{1}_{(\cdot)}$, $(\cdot)^T$, and $(\cdot)^H$  are the $\ell_2$ norm, indicator function, transpose, and conjugate transpose, respectively.  

\section{System Model and Preliminaries}
\label{sec:System Model and Preliminaries}
 This section formulates  near-field localization  and reviews necessary preliminaries. We first describe the signal model in Subsection~\ref{subsec:System Model} and  the localization problem in Subsection~\ref{subsec:Problem formulation}.  Then, we briefly review  background in  model-based and data-driven Subspace methods in Subsection~\ref{subsec: Subspace methods}. 

\subsection{Signal Model}
\label{subsec:System Model}
\begin{figure}
    \centering
    \includegraphics[width=0.8\columnwidth]{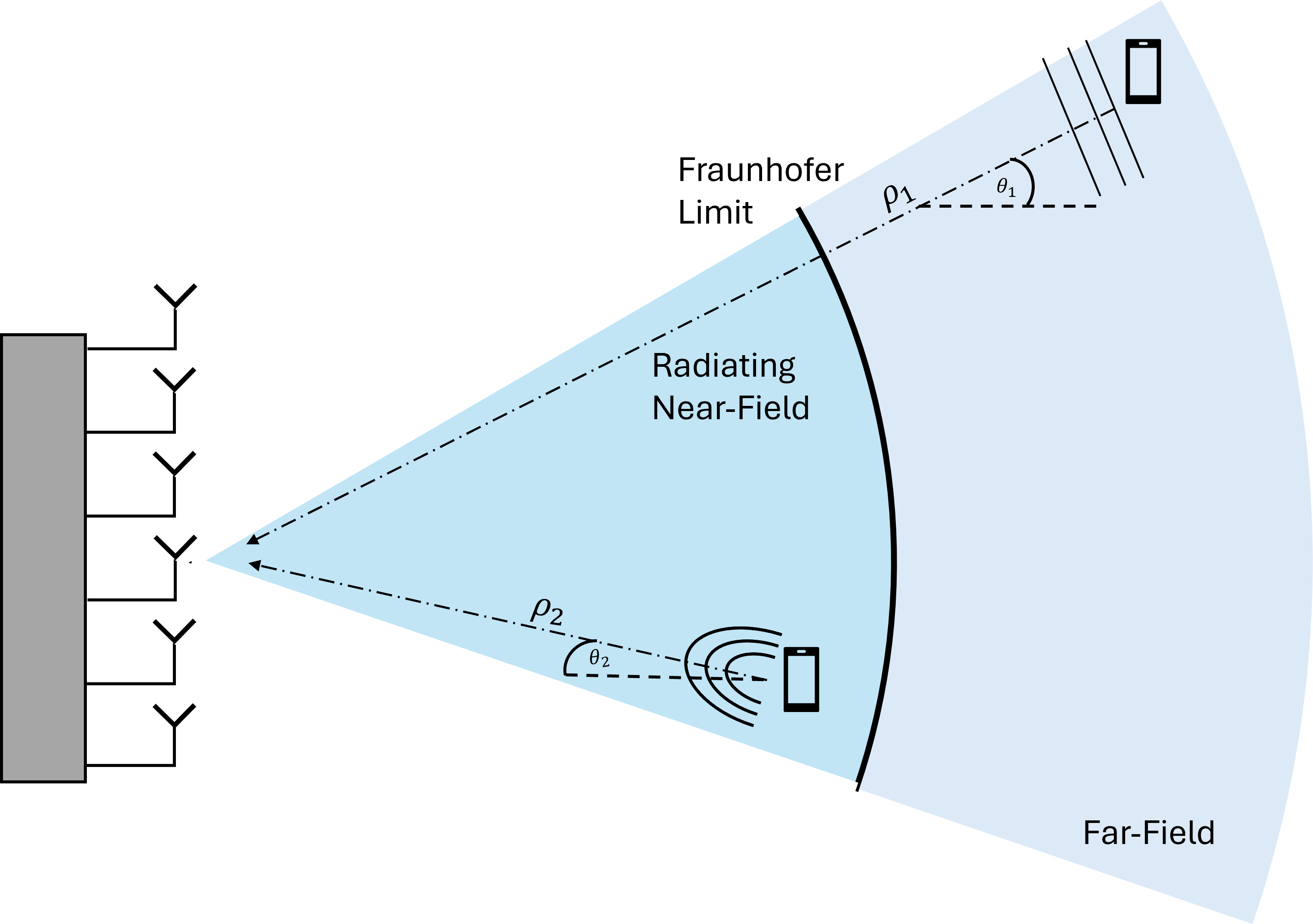}
    \caption{Multi source localization illustration.}
    \label{fig:System_Illust}
\end{figure}

The conventional modeling for passive localization in two-dimensional space considers $M$ narrowband sources impinging an \ac{ula}. The \ac{ula} has  $N>M$ elements, that are separated with spacing $d$. 
We write the  angle and distance of the $m$th source from the array as $\theta_m$ and $\rho_m$, respectively. 
We focus on settings where the sources  possibly lie in the {\em radiative near-field}~\cite{zhang20236g}, namely, the distance of the $m$th source,  $\rho_m$, can be between the Fresnel limit and the Fraunhofer limit of the array. For wavelength $\lambda$ and array aperture $D=(N-1)\cdot d$, this implies that \cite{fresnelregion2017}
\begin{equation}
\label{eq: fresnel region}
    \sqrt[3]{{D^4}/{8\,\lambda}} < \rho_m <  2 \cdot {D^{2}}/{\lambda}, \qquad \forall m\in \{1,\ldots,M\}.
\end{equation}
If the distance is larger than the Fraunhofer limit $2 \cdot {D^{2}}/{\lambda}$, then the source is said to lie in the {\em far-field}, which is the conventional regime considered in \ac{doa} estimation settings~\cite{pillai2012array}. The considered system is illustrated in Fig.~\ref{fig:System_Illust}.

The signal received at the array at time instant $t$ is described as the $N\times 1$ vector $\myVec{x}(t)$, given by:%
\begin{equation} \label{eq: single sensor in time}
    \myVec{x} (t) = \sum_{m = 1}^{M}\myVec{a}(\theta_{m}, \rho_{m})s_{m}(t) +\myVec{w}(t).
\end{equation}
In \eqref{eq: single sensor in time}, $s_m(t)$ is the $m$th source signal, $\myVec{w}(t)$ is an additive noise, and $\myVec{a}(\theta, \rho)$ is the steering vector towards $(\theta, \rho)$. 
The common approach to model the near-field steering vector, adopted in \cite{Guanghui2020, Challa1995HighOrderEsprit, 2dmusic1991,zuo2020subspace, sun2024projection, Liang2010MixFarNearLocalization, zhang2018localization, Starer1994pathfollow, zhi20071dmusicesprit}, 
writes its  $n$th element  as 
\begin{equation}
\label{eq: steering vec model}
    [\myVec{a}(\theta_{m}, \rho_{m})]_n = e^{jn\omega(\theta_{m}) + jn^{2} \psi(\theta_{m}, \rho_{m})},
\end{equation}
for $n \in\{1, \ldots, N\}$, where 
\[
\omega(\theta_{m}) \triangleq -\frac{2\pi}{\lambda} d \sin(\theta_{m}), \quad
\psi(\theta_{m}, \rho_{m}) \triangleq \frac{\pi}{\lambda} \frac{d^{2} \cos^{2}(\theta_{m})}{\rho_{m}}.
\] 
An alternative model proposed in \cite{freidlander2019} formulates the elements of the  near-field steering vector as
\begin{equation}
\label{eq: steering vec alternative model}
    [\myVec{a}(\theta_{m}, \rho_{m})]_n =\frac{\rho_{m}}{\rho_{n, m}} e^{jn\omega(\theta_{m}) + jn^{2} \psi(\theta_{m}, \rho_{m})},
\end{equation}
where $\rho_{n, m}$ is the distance of the $m$th source to the $n$th sensor element in the array. When $\rho_{m}$ is larger than the Fraunhofer limit, it holds that $\psi(\theta_{m}, \rho_{m}) \approx 0$ and $\rho_{n,m} \approx \rho_m$, and thus both \eqref{eq: steering vec model} and \eqref{eq: steering vec alternative model} reduce to the conventional \ac{ula} far-field steering vector representation~\cite{pillai2012array}.

The signal model in  \eqref{eq: single sensor in time} is defined for a single snapshot. Gathering the signals over $T$ snapshots yields a matrix signal model, which holds for both formulations of the steering vector. Letting  $\myVec{\theta} \triangleq [\theta_1,\ldots,\theta_M]^T$, 
$\myVec{\rho} \triangleq [\rho_1,\ldots,\rho_M]^T$, the matrix model is given by
\begin{equation} 
\label{eq: array signal}
    \myMat{X} = \myMat{A}(\myVec{\theta},\myVec{\rho})\myMat{S} + \myMat{W},
\end{equation}
where $\myMat{A}(\myVec{\theta},\myVec{\rho}) \in \mathbb{C}^{N \times M}$ is the steering matrix whose $m$th column is $\myVec{a}(\theta_m,\rho_m)$; while $\myMat{S} \triangleq [\myVec{s}(1),\ldots, \myVec{s}(T)]\in \mathbb{C}^{M \times T}$, 
$\myMat{X} \triangleq [\myVec{x}(1),\ldots, \myVec{x}(T)] \in \mathbb{C}^{N \times T}$,
and 
$\myMat{W} \triangleq [\myVec{w}(1),\ldots, \myVec{w}(T)]\in \mathbb{C}^{N \times T}$.

\subsection{Problem Formulation}
\label{subsec:Problem formulation}
We wish to localize the sources, namely, estimate $\myVec{\theta}$ and $\myVec{\rho}$ from the measured $\myMat{X}$. 
The number of sources $M$ is not known a-priori, but is assumed to be smaller than $N$. 
Accordingly, we are interested in an algorithm that maps $\myMat{X}$ into an estimate of the number of sources $\hat{M}\in \{0,1,\ldots,N-1\}$ as well as the corresponding estimates of the angles and ranges, respectively  denoted as
\begin{equation*}
\hat{\myVec{\theta}}\in[-\pi/2,\pi/2]^{\hat{M}}, \quad \hat{\myVec{\rho}}\in [\sqrt[3]{{D^4}/{8\,\lambda}},  2 \cdot {D^{2}}/{\lambda}]^{\hat{M}}.
\end{equation*}
Recall that for conventional far-field localization, i.e., when $\rho_m$ is larger than the Fraunhofer limit in \eqref{eq: fresnel region}, then the distance  cannot be recovered from $\myMat{X}$. Accordingly, a range of $[\hat{\myVec{\rho}}]_m= 2 \cdot {D^{2}}/{\lambda}$  estimates that the source as lying  in the far-field.

Unlike conventional approaches for localizing in the near-field~\cite{Guanghui2020, Challa1995HighOrderEsprit, 2dmusic1991,zuo2020subspace, sun2024projection, Liang2010MixFarNearLocalization, zhang2018localization, Starer1994pathfollow, zhi20071dmusicesprit}, we  focus on challenging settings, where, e.g., 
\begin{enumerate}[label={\bf C\arabic*}]
    \item \label{itm:NonCoherent} The sources are coherent, i.e., the covariance matrix of $\myVec{s}(t)$, denoted $\myMat{R}_{\myVec{S}}$, is non-diagonal. 
    \item \label{itm:MisCalib} The array includes unknown mis-calibrations, e.g., the actual element spacing may differ from the value of $d$ provided to the algorithm, and may  deviate from $\frac{\lambda}{2}$.
    \item \label{itm:SNR} The number of snapshots $T$ is small, and the \ac{snr} is low. 
\end{enumerate}
As follows from our formulation of the setup, we are also interested in coping with the following challenge:
\begin{enumerate}[label={\bf C\arabic*}, resume] 
    \item \label{itm:Mixed} Each source can be either in the far-field or in the  near-field.
\end{enumerate}

To tackle these possible issues, we assume access to labeled data, obtained from, e.g., simulations or field measurements. The data is comprised of ${J}$ pairs of observation and their  angles-ranges, denoted as
\begin{equation}
\label{eqn:Dataset}
    \mySet{D} = \big\{\big(\myMat{X}_{j}, \{(\theta_{m,j}, \rho_{m,j})\}_{m=1}^{M_{j}}\big)\big\}_{j=1}^{J}.
\end{equation}

\subsection{Model-Based and Data-Driven Subspace Methods}
\label{subsec: Subspace methods}
\subsubsection{Subspace Methods}
A leading family of algorithms for passive localization is that of {\em subspace methods}~\cite[Ch. 2]{pillai2012array}. These methods localize multiple sources in the absence of \ref{itm:NonCoherent}-\ref{itm:SNR} by extracting from the input covariance orthogonal representations of the signal and noise subspaces. In doing so, they can localize with resolution that is not dictated by the geometry of the \ac{ula}. 

Specifically, for non-coherent sources (not meeting \ref{itm:NonCoherent}), the \ac{evd} of the covariance  of $\myVec{x}(t)$ contains $M$ dominant eigenvalues, whose eigenvectors correspond to the signal subspace. The remaining $N-M$ eigenvectors, denoted by $N\times (N-M)$ matrix $\myMat{U}_{\rm W}$, represent the noise subspace and are orthogonal to the steering vectors~\cite{schmidt1986music}, i.e.,
\begin{equation}
\label{eqn:SubspaceRuule}
 \left\| \myMat{{U}}^{H}_{\rm W}\myVec{a}({\theta_{m}},{\rho_{m}})\right\|=0. \quad     \forall m \in \{1, \ldots, M\}.
\end{equation}

Subspace methods  first estimate the input covariance as
 $   \myMat{\hat{R}}_{X} = \frac{1}{T}\sum_{t=0}^{T-1}{\myVec{x}(t)\myVec{x}^{H}(t)}$,
 requiring sufficient snapshots (not meeting \ref{itm:SNR}) to estimate it. The number of sources $\hat{M}$ and noise subspace $\myMat{\hat{U}}_{W}$ are estimated from the \ac{evd} of $\myMat{\hat{R}}_{X}$,  by dividing its eigenvalues, denoted $\{\hat{\lambda}_n\}$, into  $\hat{M}$ dominant  and $N- \hat{M}$ least dominant ones (noise subspace). 
 Specifically, the setting of $\hat{M}$ is often realized via {\em thresholding}, i.e., by comparing to a fixed threshold $\thresh$ and setting
 \begin{equation}
     \hat{M} = \sum_{n=1}^{N} \myVec{1}_{\hat{\lambda}_n > \thresh}.
     \label{eqn:thresholding}
 \end{equation}

 An alternative approach treats the setting of $M$ using {\em model-order selection} tests, such as the {\em \ac{mdl}} criterion \cite{rissanen1978modeling} (which can be viewed as an approximation of the Bayesian Information criterion) or the {\em \ac{aic}}~\cite{akaike1998information}. Assuming that the signals $\myVec{s}(t)$ and the noise $\myVec{w}(t)$ are i.i.d. Gaussian, such tests estimate $M$ under the signal model in \eqref{eq: single sensor in time} as \cite{Wax1985ModelOrder} 
 \begin{align}     
     \hat{M}= &\mathop{\arg\min}_{M\in \{0,\ldots, N-1\}} -T\sum_{n = M+1}^N\log \left(\hat{\lambda}_n \right) \notag \\
     &+ T(N-M) \log\left(\frac{1}{N-M}\sum_{n=M+1}^N \hat{\lambda}_n\right)   \notag \\
     &\quad + {\frac{1}{2}\left(2M(N-M) + 1\right)} \cdot \zeta(T),
     \label{eqn:MDL}
 \end{align}
 with $\zeta(t)= \log(T)$ for \ac{mdl}, and $\zeta(T)\equiv 2$ for \ac{aic}.

 

\subsubsection{Near-Field MUSIC} 
Once $\hat{M}$ and the noise subspace  $\myMat{\hat{U}}_{W}$ are recovered, subspace methods localize the sources  based on the property in \eqref{eqn:SubspaceRuule}. When the sources lie in the far-field, and thus the \ac{ula} steering vector models linear phase accumulation, the angles $\myVec{\theta}$ can be estimated based on \eqref{eqn:SubspaceRuule} using different algorithms, including \ac{music}~\cite{schmidt1986music}, Root-\ac{music}~\cite{Barabell1983ImprovingTR}, and \ac{esprit}~\cite{roy1989esprit}.

While Root-\ac{music} and \ac{esprit} are particularly tailored for far-field sources, \ac{music} can be applied regardless of the steering matrix structure, and can thus be extended for near-field localization~\cite{2dmusic1991}.   Specifically, Near-Field \ac{music}  estimates $\myVec{\theta}$ and $\myVec{\rho}$ by finding the $\hat{M}$ peaks in the \ac{music} spectrum, given by
\begin{equation}\label{MUSIC_spectrum}
    P_{\rm MUSIC}(\theta,\rho) = \big\| {\hat{\myMat{U}}}_{W}^{H}\myVec{a}(\theta, \rho)\big\|^{-2}.
\end{equation}
 The \ac{music} spectrum in \eqref{MUSIC_spectrum} visually represents the location of the sources. It is computed by expressing the steering vectors via \eqref{eq: steering vec model} (which can be inaccurate under   \ref{itm:MisCalib}). Finding the peaks of \eqref{MUSIC_spectrum} typically requires searching over a 2D grid. 

\subsubsection{SubspaceNet}
\label{subsec: ssn}
As detailed above, subspace methods are typically studied in the context of far-field \ac{doa} recovery. There, it was recently shown in \cite{shmuel2024subspacenet} that one can incorporate deep learning techniques to make subspace methods operate in the presence of 
Challenges \ref{itm:NonCoherent}-\ref{itm:SNR}. 

Specifically, the SubspaceNet algorithm of \cite{shmuel2024subspacenet} suggested to train a \ac{dnn} to  map the observed signal $\myMat{X}$ into a {\em surrogate} covariance matrix that is useful for downstream subspace-based \ac{doa} recovery, as if \ref{itm:NonCoherent}-\ref{itm:SNR} are not present. This was achieved by, drawing inspiration from focusing methods \cite{stoica2005spectral}, first extracting the autocorrelation features as
\begin{equation}
\label{eqn:AutoCorr}
    \myMat{R}_{x}[\tau] = \frac{1}{T-\tau}\sum_{t=0}^{T-\tau}\myVec{x}(t)\myVec{x}^{H}(t+\tau),
\end{equation}
for $\tau \in {0,..., \tau_{\text{max}}}$.
  The autocorrelation features, $\{\myMat{R}_{x}[\tau]\}_{\tau=0}^{\tau_{\max}}$, are then mapped into real-valued features by concatenating the real and imaginary parts, and processing the resulting tensor through a de-noising \ac{cnn} auto-encoder with trainable parameters $\myVec{\psi}$. The network outputs are the real and imaginary parts of an $N \times N$ matrix $\myMat{K}( \myMat{X}; \myVec{\psi})$,  used to obtain a positive define surrogate covariance via
\begin{equation} \label{eq: positive define transform}
\myMat{\hat{R}}(\myMat{X};\myVec{\psi}) = \myMat{K}(\myMat{X};\myVec{\psi})\myMat{K}^H(\myMat{X};\myVec{\psi}) + \epsilon \myMat{I}_{N}, \quad \epsilon > 0.
\end{equation}

The training procedure evaluated the surrogate covariance in \eqref{eq: positive define transform} based on its usefulness for downstream subspace based \ac{doa} estimation. This was achieved by providing $\myMat{\hat{R}}(\myMat{X};\myVec{\psi})$ to a {\em differentiable} far-field subspace method (particularly \ac{esprit} and Root-\ac{music}). Leveraging the differentiability of the subsequent far-field subspace method, the internal \ac{dnn} was trained to minimize the \ac{doa} estimation error on a training dataset, where the gradients of the empirical risk were computed by backpropagating through the subspace method~\cite{shmuel2024subspacenet}.


\section{AI-Aided Subspace Methods}
\label{sec:method}
The fact that far-field \ac{ai}-augmented subspace methods rely on using a differentiable subspace method during training, and particularly \ac{esprit} or Root-\ac{music}, implies that these far-field algorithms do not extend to the radiating near-field. However, their underlying concept, advocating the usage of deep learning techniques to learn surrogate covariance matrices for subspace-based localization purposes, guides us in next proposing \ac{ai}-aided near-field subspace methods. 

Accordingly, in this section, we present two such algorithms: The first, termed {\em \name} (presented in Subsection~\ref{subsec:PMN}), extends the methodology of (far-field) SubspaceNet to Near-Field MUSIC; it does so by integrating novel tools to evaluate the resulting spectrum while facilitating accurate estimation of the number of sources. 
Our second algorithm, coined {\em \ac{dcdmusic}} (presented in Subsection~\ref{subsec:DCDMUSIC}), adopts a different approach, where instead of modifying the localization algorithm to be near-field oriented, we transform the signal covariance to correspond to far-field signals. This is achieved by decoupling angle and range recovery in a cascaded fashion, drawing inspiration from \cite{zhi20071dmusicesprit,Magoarou2019}. We conclude with a qualitative comparison and a discussion in Subsection~\ref{subsec:discussion}.



\subsection{\name}
\label{subsec:PMN}
\subsubsection{Rationale}
As discussed in Subsection~\ref{subsec: ssn}, it was shown in \cite{shmuel2024subspacenet} that the availability of data as in \eqref{eqn:Dataset} can be utilized to enable subspace methods to operate under challenges \ref{itm:NonCoherent}-\ref{itm:SNR}, by training a dedicated \ac{dnn} to map the observations in to a surrogate covariance matrix. However, directly extending this methodology to near-field scenarios (as well as to mixed near-field/far-field \ref{itm:Mixed}) is not straightforward. This limitation arises because the differentiable algorithms underpinning the training of the \ac{dnn} (\ac{esprit} or Root-MUSIC) are incompatible with the fact that the phase of the entries of the steering matrix is a non-linear function of the antenna index in the near-field. The subspace method that is suitable in the near-field, i.e., Near-Field \ac{music},  is not differentiable due to its reliance on a peak-finding operation. 

To overcome these challenges, we leverage recent developments in differentiable approximations of \ac{music}-type algorithms~\cite{chatelier2024physically}. Specifically, we allow SubspaceNet to be applied for localizing mixed near-field and far-field sources by  proposing a novel loss function, that evaluates the covariance based on two main criteria: $(i)$ its usefulness for producing a meaningful  {\em near-field \ac{music} spectrum} (as opposed to predicting the correct locations as in previous \ac{ai}-augmented subspace methods). This is achieved by encouraging the noise subspace to be as orthogonal as possible to the steering matrix at the true source locations; and $(ii)$ the {\em distinguishability of its eigenspectrum} into signal and noise subspaces. 


\subsubsection{Architecture}
Following the above rationale, we design \name~to employ a dedicated \ac{dnn} with trainable parameters $\myVec{\psi}$, whose input is the (complex-valued) autocorrelation features  $\{\myMat{R}_{x}[\tau]\}_{\tau=0}^{\tau_{\max}}$, computed via \eqref{eqn:AutoCorr}. The outputs of the \ac{dnn} are the real and imaginary parts that form the $N\times N$ matrix $\myMat{K}(\myMat{X};\myVec{\psi})$, which in turn is converted into the surrogate covariance $\hat{\myMat{R}}(\myMat{X};\myVec{\psi})$. Hence, the  \ac{dnn}  maps a $2(\tau_{\max}+1)\times N \times N$ tensor into a $2 \times N \times N$ tensor.


\begin{figure}
    \centering
    \includegraphics[width=\columnwidth]{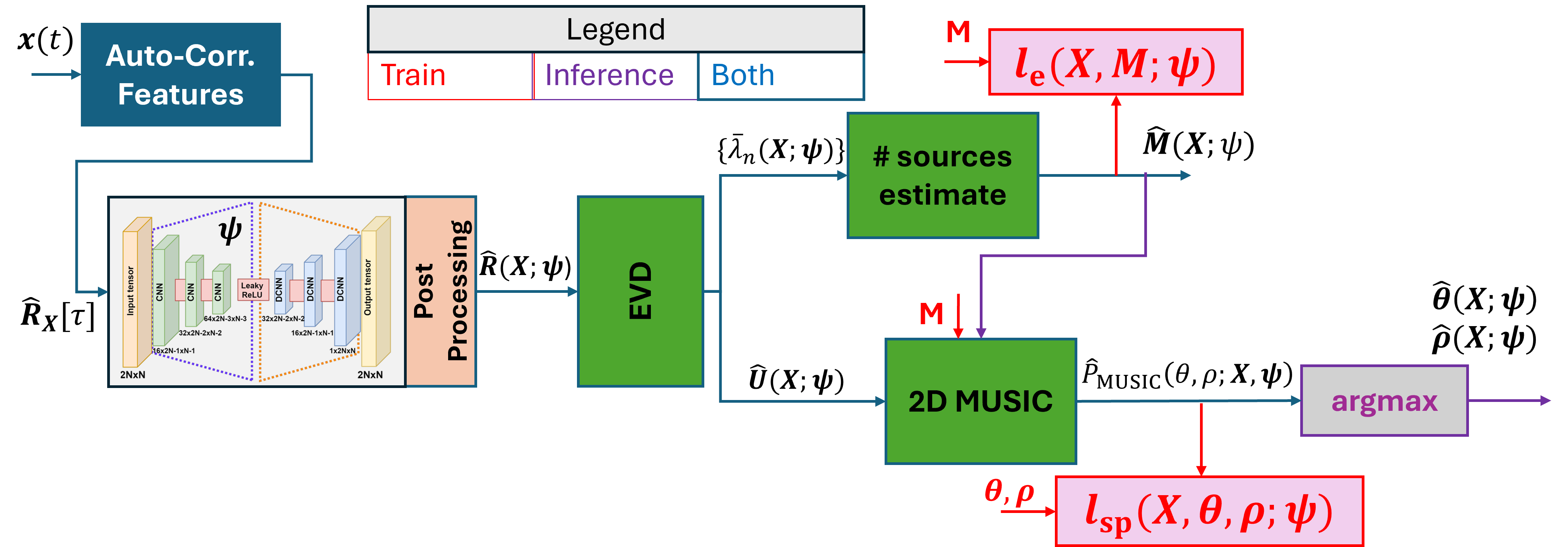}
    \caption{\name~illustration.}
    \label{fig:PeakMUSICNet_flow}
\end{figure}

We use an architecture comprised of a sequence of convolutional layers followed by a sequence of deconvolutional layers as a form of an autoencoder (with non identical input and output dimensions). Drawing inspiration from \cite{ABLE}, we employ anti-rectifier activations to avoid sparsification of the internal features while preserving the simplicity of conventional ReLU. Instead of using \eqref{eq: positive define transform} to directly obtain the surrogate covariance as in \cite{shmuel2024subspacenet}, we introduce a normalization layer whose purpose is to regularize the variation in the eigenvalues, that play a key role in identifying the number of sources. To that aim, we normalize \eqref{eq: positive define transform} by its largest eigenvalue, i.e., its  \textit{spectral norm}, and thus the surrogate covariance is obtained as
\begin{equation} \label{eq: positive define transform_upd}
\myMat{\hat{R}}(\myMat{X};\myVec{\psi}) = \frac{\myMat{K}(\myMat{X};\myVec{\psi})\myMat{K}^H(\myMat{X};\myVec{\psi}) + \epsilon \myMat{I}_{N}}{\|\myMat{K}(\myMat{X};\myVec{\psi})\myMat{K}^H(\myMat{X};\myVec{\psi}) + \epsilon \myMat{I}_{N}\|_2}.
\end{equation}
The last layer of the architecture is a learnable skip connection with a learnable parameter $\alpha$, which is constrained to the range $(0, 1)$. This parameter introduces a correction term to the empirical covariance of the input signal, $\myMat{R}_{x}[\tau = 0]$, by defining the surrogate covariance as follows: 
\[
\myMat{\hat{R}}(\myMat{X}; \myVec{\psi}) = \myMat{\hat{R}}(\myMat{X}; \myVec{\psi}) + \alpha \myMat{R}_{x}[\tau = 0].
\]

During inference, the surrogate covariance matrix is processed with Near-Field \ac{music}  for angle and range estimation. 
First, the eignenvalues of $\hat{\myMat{R}}(\myMat{X};\myVec{\psi})$ are used to identify the number of sources $\hat{M}$ via either \eqref{eqn:thresholding} or \eqref{eqn:MDL}.
    Next, the $N\times \hat{M}$ noise subspace  is computed from  $\hat{\myMat{R}}(\myMat{X};\myVec{\psi})$, and is used to formulate the \ac{music} spectrum, denoted $P_{\text{MUSIC}}(\theta, \rho; \myMat{X}, \myVec{\psi})$, using \eqref{MUSIC_spectrum}.
    Then, the locations are recovered from the $\hat{M}$ peaks of $P_{\text{MUSIC}}(\theta, \rho; \myMat{X}, \myVec{\psi})$. Sources located at the  upper limit of the $\rho$ range  are labeled as far-field sources.
The  procedure is summarized as Algorithm~\ref{alg:pmn_inference}, and illustrated in Fig.~\ref{fig:PeakMUSICNet_flow}.

\begin{algorithm}
\caption{\name~Localization}
\label{alg:pmn_inference}
\SetKwInOut{Initialization}{Init}
\Initialization{Trained \ac{dnn} $\myVec{\psi}$; hyperparameters $(\epsilon, \tau_{\max})$}
\SetKwInOut{Input}{Input} 
\Input{Observations $\myMat{X}$}
    Compute $\{\myMat{R}_{x}[\tau]\}_{\tau=0}^{\tau_{\max}}$ via \eqref{eqn:AutoCorr}\;
    Apply \ac{dnn} and \eqref{eq: positive define transform_upd} to get $\hat{\myMat{R}}(\myMat{X};\myVec{\psi})$\;
    Set  $\{\bar{\lambda}_n(\myMat{X};\myVec{\psi})\}$ as eigenvalues of $\hat{\myMat{R}}(\myMat{X};\myVec{\psi})$\; 
    Compute $\hat{M}$ from  $\{\bar{\lambda}_n(\myMat{X};\myVec{\psi})\}$  via \eqref{eqn:thresholding} or  \eqref{eqn:MDL}\;
    Formulate spectrum $P_{\text{MUSIC}}(\theta, \rho; \myMat{X}, \myVec{\psi})$, using \eqref{MUSIC_spectrum}\;
    Set $\hat{\myVec{\theta}}, \hat{\myVec{\rho}}$ as $\hat{M}$ peaks of $P_{\text{MUSIC}}(\theta, \rho; \myMat{X}, \myVec{\psi})$\;
\KwRet{$\hat{\myVec{\theta}}, \hat{\myVec{\rho}}$}
\end{algorithm}



\subsubsection{Training} \label{subsubsec: nfssn train}
The learning procedure, i.e., the setting of the trainable parameters $\myVec{\psi}$ from the dataset $\mySet{D}$, aims at having \name~produce accurate predicted locations based on the  signals in $\mySet{D}$. While presumably one would wish to do so by converting \name~into a trainable discriminative machine learning model~\cite{shlezinger2022discriminative}, following the conventional approach in model-based deep learning~\cite{shlezinger2023model}, such an approach cannot be carried out directly, as gradients cannot propagate through the near-field \ac{music} algorithm. To overcome this challenge we introduce a novel loss function that $(i)$ evaluates the near-field \ac{music} spectrum {\em prior to peak finding}; and $(ii)$ encourages accurate recovery of the number of sources via  thresholding or model-order selection tests. 

{\bf Spectrum Loss:}
To evaluate localization accuracy, we formulate a  loss  inspired by the core principles of subspace methods.
Instead of  computing the loss based on the predicted source locations, we seek orthogonality of the noise subspace to the steering matrix. This is achieved by encouraging the surrogate covariance to produce a near-field \ac{music} spectrum whose inverse is minimized at the sources' locations, namely,
\begin{equation}
    \label{eq: music spectrum loss}
    l_{\text{sp}}(\myMat{X}, \myVec{\theta}, \myVec{\rho}; \myVec{\psi}) = \sum_{m=1}^{M} P_{\text{MUSIC}}^{-1}(\theta_{m}, \rho_{m}; \myMat{X}, \myVec{\psi}).
\end{equation}
By minimizing \eqref{eq: music spectrum loss}, the surrogate covariance matrix is optimized to yield a noise subspace that is highly orthogonal to the steering matrix at the source positions.

{\bf Model-Order Loss:} 
We design a  regularization component  to foster a more distinct eigenspectrum, and particularly, enhance the ability to recover $M$ from the (sorted) eigenvalues of the surrogate covariance $\myMat{\hat{R}}(\myMat{X};\myVec{\psi})$, denoted by  $\{\bar{\lambda}_n(\myMat{X};\myVec{\psi})\}_{n=1}^N$. In order to tailor the loss term to the downstream method for detecting $M$, we propose either of the following formulations:
\begin{itemize}
    \item {\em Thresholding:} 
    When the subspace method recovers $M$ via \eqref{eqn:thresholding}, we propose to penalize the  weights $\myVec{\psi}$ to yield an eigenspectrum where the $M$th eigenvalue is misclassified as noise or the $M+1$th eigenvalue is misclassified as a signal.
    The proposed regularization term is
    \begin{equation} \label{eq:eigenRegularization}
       l_{\rm e}(\myMat{X},M;\myVec{\psi}) = \prod\nolimits_{m=M}^{M+1}\left(\bar{\lambda}_m(\myMat{X};\myVec{\psi})- \thresh\right).
    \end{equation} 
     $\thresh$ here is treated as a hyperparameter.
    \item {\em Model-Order Selection Test:} 
    When $M$ is recovered during inference using a model-order selecton test, e.g., \ac{mdl} or \ac{aic} as in \eqref{eqn:MDL}, we use the explicit test as a loss term, encouraging the model to minimize it at the true number of sources. Accordingly, the regularization term is comprised of the terms \eqref{eqn:MDL} that depend on $\myVec{\psi}_{\rm a}$, i.e., 
    \begin{align} 
       &l_{\rm e}(\myMat{X},M;\myVec{\psi}) = \frac{-1}{N-M}\sum_{n = M+1}^N\log \left(\bar{\lambda}_n(\myMat{X};\myVec{\psi}) \right) \notag \\
     &\qquad + \log\left(\frac{1}{N-M}\sum_{n=M+1}^N \bar{\lambda}_n(\myMat{X};\myVec{\psi})\right).   
     \label{eq:ModelOrderRegularization}
    \end{align} 
\end{itemize}

{\bf Training Algorithm:}
The overall loss function used for training \name~based on a dataset $\mySet{D}$ is given by
\begin{equation}
    \label{eq: loss over dataset for musicpeaknet}
    \mathcal{L}_{\mySet{D}}(\myVec{\psi}) \!=\! \frac{1}{|\mySet{D}|} \sum_{j=1}^{|\mySet{D}|} l_{\text{sp}}(\myMat{X}_{j}, \myVec{\theta}_{j}, \myVec{\rho}_{j}; \myVec{\psi}) \!+\!\mu_{e}l_{\rm e}(\myMat{X}_{j},M_j;\myVec{\psi}),
\end{equation}
where $\mu_{\text{e}} > 0$ is a regularization coefficient, balancing the spectrum loss with the model-order loss.
The selection of \eqref{eq: music spectrum loss} allows  training Algorithm~\ref{alg:pmn_training} as a machine learning architecture using conventional first-order optimizers,  since it enables gradient propagation through each step of the algorithm (see Fig.~\ref{fig:PeakMUSICNet_flow}). The complete processing chain from the empirical autocorrelation features to the near-field \ac{music} spectrum, which includes traditional \ac{cnn} layers, as well as differentiable operations like \ac{evd} (see, e.g., \cite{solomon2019deep}) and tensor multiplications, is end-to-end differentiable. 
The  training procedure, based on mini-batch \ac{sgd}, is summarized in Algorithm~\ref{alg:pmn_training}. 

\begin{algorithm}
\caption{\ac{sgd} Training of \name}
\label{alg:pmn_training}
\SetKwInOut{Initialization}{Init}
\Initialization{Learning parameters  ($\mu, e_{\max}, B$), hyperparameters $(\epsilon, \tau_{\max})$, initial weights $\myVec{\psi}$;}
\SetKwInOut{Input}{Input} 
\Input{Training dataset $\mySet{D}$}
\For{$e=1,\ldots, e_{\max}$}
{
    Randomly divide $\mySet{D}$ into $B$ batches\;
    \For{each batch $\mySet{D}^{(b)}$}
    {
        Compute   $\myMat{\hat{R}}(\myMat{X}_j^{(b)}; \myVec{\psi})$, $\forall \myMat{X}_j^{(b)} \in \mySet{D}^{(b)}$\;
        Compute loss $\mathcal{L}_{\mySet{D}^{(b)}}(\myVec{\psi})$ using \eqref{eq: loss over dataset for musicpeaknet}\;
        Update parameters $\myVec{\psi} \gets \myVec{\psi} - \mu\nabla_{\myVec{\psi}} \mathcal{L}_{\mySet{D}^{(b)}}(\myVec{\psi})$\;
    }
} 
\KwRet{$\myVec{\psi}$}
\end{algorithm}

\subsection{DCD-MUSIC}
\label{subsec:DCDMUSIC}
\subsubsection{Rationale}
The formulation of \name~in Subsection~\ref{subsec:PMN} aims at modifying existing far-field \ac{ai}-aided subspace methods (and particularly SubspaceNet), to be applicable with near-field subspace methods. In doing so, \name~inherits some of the shortcomings of these classical methods, e.g., the need to apply a 2D search over the \ac{music} spectrum during inference, whose granularity limits resolution. 
Here, we derive an alternative scheme, coined \ac{dcdmusic}, that follows an opposite line-of-thought: instead of modifying the subspace method from far-field to near-field, as in \name, it {\em recasts the signal statistics as originating from far-field sources}, with which far-field \ac{ai}-aided subspace methods can be applied. 

Specifically, the design of \ac{dcdmusic} is motivated by two key observations: 
$(i)$ 2D \ac{music} variants (e.g., Near-Field \ac{music}) can often be approached, with reduced complexity, by a cascade of \ac{esprit} and one-dimensional \ac{music}, each applied along a distinct search coordinate~\cite{zhi20071dmusicesprit,Magoarou2019}; 
 $(ii)$ as  demonstrated in~\cite{shmuel2024subspacenet}, \ac{dnn}-based covariance pre-processing can enable interpretable far-field subspace methods to recover \ac{doa} in scenarios where signals deviate from ideal far-field models.
Building on these insights, \ac{dcdmusic} performs near-field localization using a two-stage process, which first produces a far-field surrogate covariance for angle recovery, after which the range for each source is separately recovered. 

\subsubsection{Architecture} 
\begin{figure}
    \centering
  \includegraphics[width=\columnwidth]{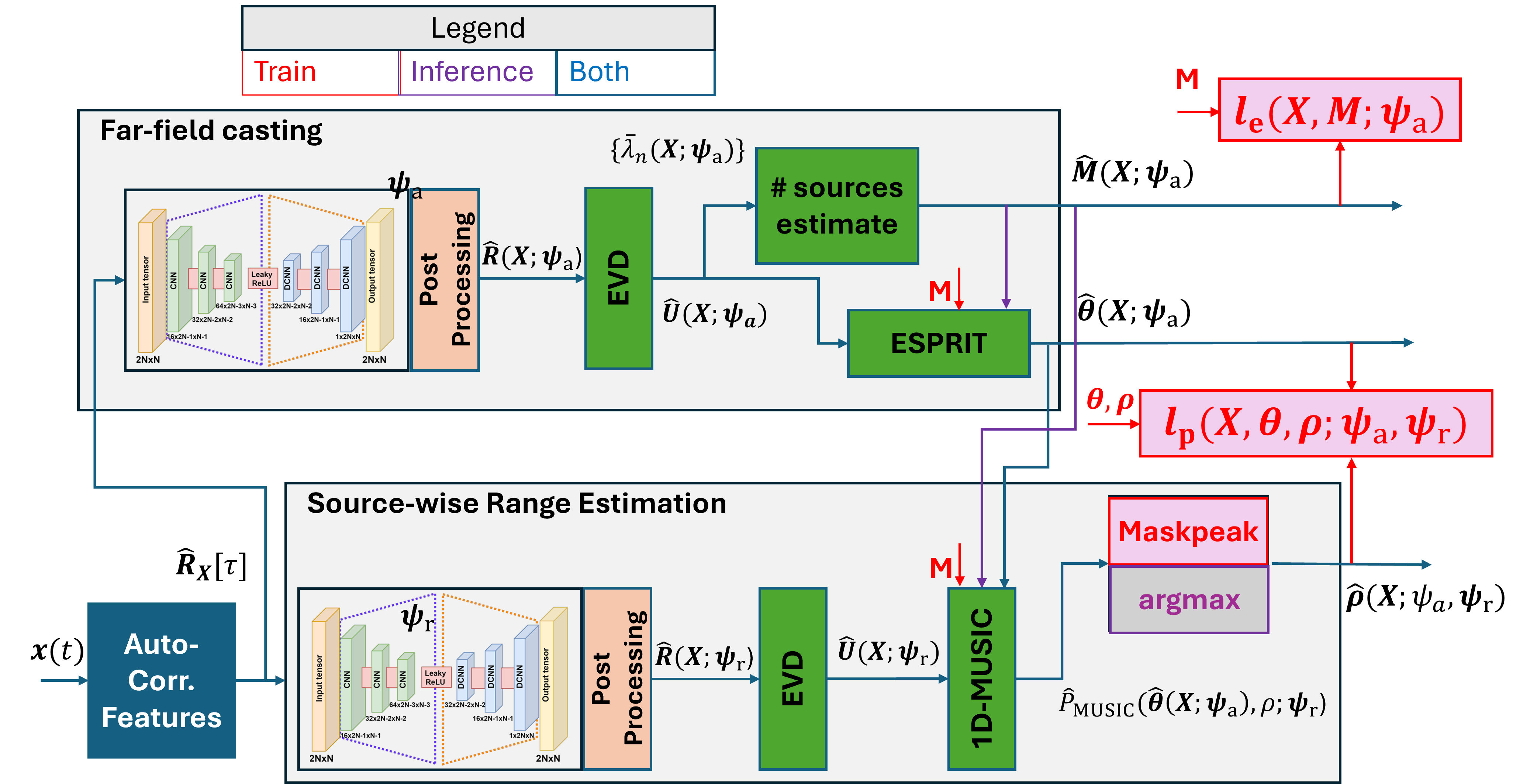}
    \caption{\ac{dcdmusic} illustration.}
    \label{fig:DCDMUSIC_flow}
\end{figure}

As outlined above, \ac{dcdmusic} is comprised of two modules: {\em Far-Field Casting},  and {\em Source-Wise Range Estimate}. We next elaborate on each of these stages, with the  overall two-stage process is depicted Fig.~\ref{fig:DCDMUSIC_flow}.

{\bf Far-Field Casting:}
First, a \ac{dnn}-based subspace-oriented surrogate covariance extraction transforms the  signal into a far-field representation. The far-field surrogate covariance is used to estimate number of sources, $M$, and their angles, $\myVec{\theta}$.

Specifically, we employ the \ac{dnn} architecture detailed in Subsection~\ref{subsec:PMN} with parameters $\myVec{\psi}_{\rm a}$. The \ac{dnn} learns to map the input $\myMat{X}$ into a surrogate covariance denoted $\hat{\myMat{R}}(\myMat{X};\myVec{\psi}_{\rm a})$, that is amenable to far-field subspace extraction of $M$ and $\myVec{\theta}$. The former is carried out using either thresholding \eqref{eqn:thresholding} or model-order selection tests \eqref{eqn:MDL} (neither of which was included in the formulation of SubspaceNet in \cite{shmuel2024subspacenet}, and is adopted from \name).   The angles of the sources are estimated from the covariance as $\hat{\myVec{\theta}}(\myMat{X};\myVec{\psi}_{\rm a})$  using \ac{esprit}. 


{\bf Source-Wise Range Estimate:} 
Having   estimated  $M$  as 
 $\hat{M}(\myMat{X};\myVec{\psi}_{\rm a})$ and $\myVec{\theta}$ as $\hat{\myVec{\theta}}(\myMat{X};\myVec{\psi}_{\rm a})$,   we proceed to recover $\myVec{\rho}$. This is achieved by searching the \ac{music} spectrum for a peak for each estimated $\hat{\theta}_m(\myMat{X};\myVec{\psi}_{\rm a})$. Since we wish to cope with \ref{itm:NonCoherent}-\ref{itm:SNR}, we cannot directly apply \ac{music} to the empirical covariance, and should
again seek a surrogate covariance using a \ac{dnn}. 
Unlike the \ac{dnn}  $\myVec{\psi}_{\rm a}$, here the surrogate covariance is not required to represent far-field signaling. Thus, a dedicated \ac{cnn} autoencoder with parameters $\myVec{\psi}_{\rm r}$  is used. Its output, denoted $\myMat{\hat{R}}(\myMat{X};\myVec{\psi}_{\rm r})$, is obtained using similar processing to those in \eqref{eq: positive define transform_upd}. 

The surrogate covariance is then used $\hat{M}(\myMat{X};\myVec{\psi}_{\rm a})$ times, for estimating the ranges of the sources for each of the estimated angles using one-dimensional \ac{music}.
Accordingly, by letting $\hat{\myMat{U}}_W(\myMat{X};\myVec{\psi}_{\rm r}, \myVec{\psi}_{\rm a})$ be the $N - \hat{M}(\myMat{X};\myVec{\psi}_{\rm a})$ least dominant eigenvectors of $\myMat{\hat{R}}(\myMat{X};\myVec{\psi}_{\rm r})$, we estimate  $\hat{\rho}_m(\myMat{X};\myVec{\psi}_{\rm r}, \myVec{\psi}_{\rm a})$ via
\begin{align}
    &{\arg\max}_{\rho} {\big\| {\hat{\myMat{U}}}_{W}^{H}(\myMat{X};\myVec{\psi}_{\rm r}, \myVec{\psi}_{\rm a})\myVec{a}(\hat{\theta}_m(\myMat{X};\myVec{\psi}_{\rm a}), \rho)\big\|^{-2}},
    \label{eqn:estRho}
\end{align}
for each $m \in \{1,\ldots, \hat{M}(\myMat{X};\myVec{\psi}_{\rm a})\}$. The $\arg \max$ operation in \eqref{eqn:estRho} is computed by grid search over a range holding \eqref{eq: fresnel region}. Sources  at the  far edge of the range, i.e., at the  limit, are labeled as far-field sources.
The resulting \ac{dcdmusic} algorithm is summarized as Algorithm~\ref{alg:Inference algorithm}.
{The chosen configuration employs \ac{esprit} for angle estimation followed by \ac{music}  for range recovery. While alternative designs are possible, such as using \acl{rmusic} for angle estimation combined with a spectrum-based loss for range, instead of approximation of the peak finding operation, we opt the ESPRIT-MUSIC configuration with maskpeak based on empirical trials across our evaluated scenarios.
}

\begin{algorithm}
    \caption{\acs{dcdmusic} Localization}
    \label{alg:Inference algorithm} 
    \SetKwInOut{Initialization}{Init}
    \Initialization{Trained  \acp{dnn}  $\myVec{\psi}_{\rm r}, \myVec{\psi}_{\rm a}$; hyperparameters $(\epsilon, \tau_{\max})$.}
    \SetKwInOut{Input}{Input} 
    \Input{Observations $\myMat{X}$;}
    {    
      Apply \acp{dnn} to  $\myMat{X}$ to obtain $\hat{\myMat{R}}(\myMat{X};\myVec{\psi}_{\rm a})$, $\hat{\myMat{R}}(\myMat{X};\myVec{\psi}_{\rm r})$\;
    Set  $\{\bar{\lambda}_n(\myMat{X};\myVec{\psi}_{\rm a})\}$ as eigenvalues of $\hat{\myMat{R}}(\myMat{X};\myVec{\psi}_{\rm a})$\; 
    Compute $\hat{M}$ from  $\{\bar{\lambda}_n(\myMat{X};\myVec{\psi}_{\rm a})\}$  via \eqref{eqn:thresholding} or  \eqref{eqn:MDL}\;
      Estimate $\hat{\myVec{\theta}}$  from $\hat{\myMat{R}}(\myMat{X};\myVec{\psi}_{\rm a})$ via far-field \acs{esprit}\;
      \For{$m=1,\ldots, \hat{M}$}
      {
        Estimate $\hat{\rho}_m$ via \eqref{eqn:estRho};
      } 
    }\KwRet{$\hat{\myVec{\theta}}, \hat{\myVec{\rho}}$}
\end{algorithm}

\subsubsection{Training}
We train \ac{dcdmusic}, namely, set $\myVec{\psi}_{\rm a}, \myVec{\psi}_{\rm r}$ from the data $\mySet{D}$ in \eqref{eqn:Dataset}, 
in three sequential steps, designed to gradually tune each of the modules of \ac{dcdmusic}. These training stages are termed {\em Angle},  {\em Range}, and {\em Position Training}.

\textbf{Angle Training:} 
We first train only $\myVec{\psi}_{\rm a}$ by evaluating  $\hat{\myMat{R}}(\myMat{X};\myVec{\psi}_{\rm a})$ based on $(i)$ its usefulness for angle recovery via the (differentiable) far-field \acs{esprit}; and $(ii)$ the accuracy of the recovered number of sources.  

To evaluate angle recovery, we use use the \ac{rmspe} loss, which accounts for the periodic nature of the angles \cite{routtenberg2011bayesian}. This loss term is given by
\begin{align}
   \!\! l(\myMat{X},\myVec{\theta};\myVec{\psi}) \!=\! \min_{\myMat{P} \in \mySet{P}_M}\frac{ 
    \big\| {\rm mod}_{\pi}\big(\myVec{\theta} 
   \!-\! \myMat{P}\hat{\myVec{\theta}}\big(\hat{\myMat{R}}(\myMat{X};\myVec{\psi})\big)\big)\big\|}{\sqrt{M}},
   \label{eqn:RMSPE}
\end{align}
whereas $\mySet{P}_M$ be the set $M\times M$ permutation matrices, and  ${\rm mod}_{\pi}$ denote  modulo $\pi$. 
To encourage accurate recovery of $M$, we employ the model-order loss detailed in Subsection~\ref{subsec:PMN}. Thus, the overall loss term used for training $\myVec{\psi}_{\rm a}$   over the dataset $\mySet{D}$ is 
\begin{equation} \label{eq:angle training loss over dataset}
    \mathcal{L}_{\mySet{D}}^{\rm a}(\myVec{\psi}_{\rm a}) \!=\! \frac{1}{|\mySet{D}|}\sum_{j=1}^{|\mySet{D}|}
    l_{\rm a}(\myMat{X}_{j},\myVec{\theta}_{j};\myVec{\psi}_{\rm a})  + \mu_{\text{e}}  l_{\rm e}(\myMat{X}_{j},M_{j};\myVec{\psi}_{\rm a}),
\end{equation}
where $\mu_{\text{e}} > 0$ is a regularization coefficient.

\textbf{Range Training:} 
The second step trains $\myVec{\psi}_{\rm r}$ using the input and the angles $\hat{\myVec{\theta}}$ recovered by  the trained  $\myVec{\psi}_{\rm a}$. Unlike the estimation of far-field angles, range recovery cannot be carried out using differentiable subspace methods (\ac{esprit} or Root-\ac{music}) even if none of \ref{itm:NonCoherent}-\ref{itm:Mixed} hold, due to its  nonlinear phase accumulation. 
Accordingly, we apply one-dimensional \ac{music} for each angle $\hat{\theta}_m(\myMat{X};\myVec{\psi}_{\rm a})$, using the surrogate covariance $\hat{\myMat{R}}(\myMat{X};\myVec{\psi}_{\rm r})$ and the steering vectors $\myVec{a}(\hat{\theta}_m(\myMat{X};\myVec{\psi}_{\rm a}), \rho)$, to estimate the distance as $\hat{\rho}_m\big(\hat{\myMat{R}}(\myMat{X};\myVec{\psi}_{\rm r}), \hat{\myVec{\theta}}(\myMat{X};\myVec{\psi}_{\rm a})\big)$. As \ac{music} is applied here each time only to recover a single range, we can use its output as part of our loss function (instead of monitoring the spectrum directly as we did when recovering multiple source in Subsection~\ref{subsec:PMN}), using a differentiable approximation detailed below in the overall training algorithm.

Since each distance is associated with a specific angle, the ordering  in $\hat{\myVec{{\rho}}}$ matches that in $\hat{\myVec{{\theta}}}$.
Consequently, the loss term here for an input $\myMat{X}$ with distances $\myVec{\rho}$ is 
\begin{equation} \label{eq:range training loss}
    \!\! l_{\rm r}(\myMat{X},\myVec{\rho};\myVec{\psi}_{\rm r}) \!=\! \frac{ 
    \big\| \myVec{\rho} 
   \!-\! \myMat{P}_{\theta}\hat{\myVec{\rho}}\big(\hat{\myMat{R}}(\myMat{X};\myVec{\psi}_{\rm r}), \hat{\myVec{\theta}}(\myMat{X};\myVec{\psi}_{\rm a})\big)\big\|}{\sqrt{M}},
\end{equation}
where $\myMat{P}_{\theta}$ is the angles permutation from \eqref{eqn:RMSPE}. For far-field sources,  the  distance is the farthest edge of the search range. The overall loss over the dataset $\mySet{D}$ is
\begin{equation} \label{eq:range training loss over dataset}
    \mathcal{L}_{\mySet{D}}^{\rm r}(\myVec{\psi}_{\rm r}) \!=\! \frac{1}{|\mySet{D}|}\sum_{j=1}^{|\mySet{D}|}
    l_{\rm r}(\myMat{X}_{j},\myVec{\rho}_{j};\myVec{\psi}_{\rm r}).
\end{equation}

As an initial warm-up stage, we begin training by using the ground-truth angles rather than the angles recovered from the trained $\myVec{\psi}_{\rm a}$. This approach mitigates the effect of noisy angle estimates during the early stages of range recovery. After a few epochs, once the model has learned to accurately estimate the ranges, we switch to using the predicted angles for the remainder of the training process to adapt our model to yield better ranges estimation even for noisy angles.

\textbf{Position Training:} 
We conclude by jointly tuning $\myVec{\psi}_{\rm r}$ and $\myVec{\psi}_{\rm a}$ based on the  Cartesian position. Let $\myMat{C}(\myVec{\theta}, \myVec{\rho})$ be the $M\times 2$ Cartesian representation of $\myVec{\theta}, \myVec{\rho}$. The resulting position loss is
\begin{equation} \label{eq: RMSPE_cartesian_coordinates}
    \!\! l_{\rm p}(\myMat{X}, \myVec{\theta}, \myVec{\rho};\myVec{\psi}_{\rm a}, \myVec{\psi}_{\rm r}) \!=\!
    \min_{\myMat{P} \in \mathcal{P}_M} \frac{\big\|\myMat{C}(\myVec{\theta}, \myVec{\rho}) - \myMat{P}\myMat{C}(\hat{\myVec{\theta}}, \hat{\myVec{\rho}}) \big\|}{\sqrt{M}}.
\end{equation}
The loss over  $\mySet{D}$ used in the final training stage is
\begin{align} \label{eq:position training loss}
    \mathcal{L}_{\mySet{D}}^{\rm p}(\myVec{\psi}_{\rm a}, \myVec{\psi}_{\rm r}) = \frac{1}{|\mySet{D}|}\sum_{j=1}^{|\mySet{D}|}&
    l_{\rm p}(\myMat{X_{j}},\myVec{\theta}_{j},\myVec{\rho}_{j};\myVec{\psi}_{\rm a}, \myVec{\psi}_{\rm r})  \notag \\
    &+ \mu_{\text{e}}  l_{\rm e}(\myMat{X}_{j},\myVec{\theta}_{j};\myVec{\psi}_{\rm a}).
\end{align}

{\bf Training Algorithm:}
We train \ac{dcdmusic}  using standard deep learning methods based on \ac{sgd} in the following stages: 
\begin{enumerate}[label={S\arabic*}] 
    \item  Train $\myVec{\psi}_{\rm a}$ based on $\mathcal{L}_{\mySet{D}}^{\rm a}(\myVec{\psi}_{\rm a})$;
    \item  Train $\myVec{\psi}_{\rm r}$ based on $\mathcal{L}_{\mySet{D}}^{\rm r}(\myVec{\psi}_{\rm r})$; 
    \item Further tune  $\myVec{\psi}_{\rm a}$ and $\myVec{\psi}_{\rm r}$ based on $\mathcal{L}_{\mySet{D}}^{\rm p}(\myVec{\psi}_{\rm a}, \myVec{\psi}_{\rm r})$. 
\end{enumerate}

A core challenge with this procedure stems from the fact that recovering ranges via \ac{music}, i.e., ${\arg \max}$ of \eqref{MUSIC_spectrum}, is non-differentiable, limiting the usage of \ac{sgd} in stages $(ii)$ and $(iii)$. 
However, as \ac{dcdmusic} employs one-dimensional \ac{music} for recovering a single range each time, its non-differentiability can be faithfully approached using a differentiable mapping. Specifically, drawing inspiration from  \cite{mateos2023model}, we  approximate ${\arg \max}$ with a differentiable {\em Maskpeak} when taking the gradients. Let $L$ be the mask size. We compute the gradients approximating $\hat{\rho}_m^{\rm peak} =\arg \max_{\rho} P_{\rm MUSIC}(\hat{\theta}_m, \rho)$ by observing a window denoted $\myVec{\rho}_m^{\rm mask}$ over the search range of size $2L+1$ centered at $\hat{\rho}_m^{\rm peak}$. Then we take the gradients over the weighted sum of the masked ranges dictionary by, 
\begin{equation}
    \hat{\rho}_m = (\myVec{\rho}_m^{\rm mask})^T {\rm softmax}\left( P_{\rm MUSIC}(\hat{\theta}_m, \myVec{\rho}_m^{\rm mask})\right),
    \label{eqn:Maskpeak}
\end{equation}
with $P_{\rm MUSIC}(\cdot)$ computed using $\hat{\myMat{R}}(\myMat{X};\myVec{\psi}_{\rm r})$. To sharpen the approximation, $L$ gradually decreases each epoch. We summarize the  training method using mini-batch \ac{sgd} as Algorithm~\ref{alg:dcd_training}.

\begin{algorithm}
\caption{\ac{sgd} Training of \ac{dcdmusic}}
\label{alg:dcd_training}
\SetKwInOut{Initialization}{Init}
\Initialization{Learning parameters  $(\mu, e_{\max}, B, \{L_e\})$, hyperparameters $(\epsilon, \tau_{\text{max}})$, initial weights $\myVec{\psi}_{\rm a}, \myVec{\psi}_{\rm r}$;}
\SetKwInOut{Input}{Input} 
\Input{Training dataset $\mySet{D}$}

\nonl\textbf{Stage 1: Angle Training}\;
\For{$e=1,\ldots, e_{\max}$}
{
    Randomly divide $\mySet{D}$ into $B$ batches\;
    \For{each batch $\mySet{D}^{(b)}$}
    {
        Compute   $\myMat{\hat{R}}(\myMat{X}_j^{(b)}; \myVec{\psi}_{\rm a})$, $\forall \myMat{X}_j^{(b)} \in \mySet{D}^{(b)}$\;
        Estimate angles using \ac{esprit}\;
        Compute loss $\mathcal{L}^{\rm a}_{\mySet{D}^{(b)}}(\myVec{\psi}_{\rm a})$ using \eqref{eq:angle training loss over dataset}\;
        Update  $\myVec{\psi}_{\rm a} \gets \myVec{\psi}_{\rm a} - \mu\nabla_{\myVec{\psi}_{\rm a}} \mathcal{L}_{\mySet{D}^{(b)}}(\myVec{\psi}_{\rm a})$\;
    }
} 

\nonl\textbf{Stage 2: Range Training}\;
\For{$e=1,\ldots, e_{\text{max}}$}
{
    Randomly divide $\mySet{D}$ into $B$ batches\;
    \For{each batch $\mySet{D}^{(b)}$}
    {
        Compute   $\myMat{\hat{R}}(\myMat{X}_j^{(b)}; \myVec{\psi}_{\rm r})$, $\forall \myMat{X}_j^{(b)} \in \mySet{D}^{(b)}$\;
        Estimate ranges using \eqref{eqn:Maskpeak} with mask $L_e$\;
        Compute loss $\mathcal{L}^{\rm r}_{\mySet{D}^{(b)}}(\myVec{\psi}_{\rm r})$ using \eqref{eq:range training loss over dataset}\;
        Update  $\myVec{\psi}_{\rm r} \gets \myVec{\psi}_{\rm r} - \mu\nabla_{\myVec{\psi}_{\rm r}} \mathcal{L}_{\mySet{D}^{(b)}}(\myVec{\psi}_{\rm r})$\;
    }
} 

\nonl\textbf{Stage 3: Position Training}\;
\For{$e=1,\ldots, e_{\text{max}}$}
{
    Randomly divide $\mySet{D}$ into $B$ batches\;
    \For{each batch $\mySet{D}^{(b)}$}
    {
        Apply Algorithm~\ref{alg:Inference algorithm} with parameters $\myVec{\psi}=(\myVec{\psi}_{\rm a},\myVec{\psi}_{\rm r})$ to batch while replacing \eqref{eqn:estRho} with \eqref{eqn:Maskpeak} of mask $L_e$\; 
        Compute loss $\mathcal{L}^{\rm p}_{\mySet{D}^{(b)}}(\myVec{\psi}_{\rm a},\myVec{\psi}_{\rm r})$ using \eqref{eq:position training loss}\;
        Update  $ \myVec{\psi}\gets\myVec{\psi} - \mu\nabla_{\myVec{\psi}} \mathcal{L}_{\mySet{D}^{(b)}}(\myVec{\psi})$\;
    }
}

\KwRet{$\myVec{\psi}_{\rm a},\myVec{\psi}_{\rm r}$}
\end{algorithm}

\subsection{Discussion}
\label{subsec:discussion}

We introduce two AI-aided subspace methods, \name~and \ac{dcdmusic}. Both algorithms are designed to preserve the interpretable operation of subspace-based localization, while enabling it to deal with challenges \ref{itm:NonCoherent}-\ref{itm:Mixed} via \ac{ai} augmentation. To understand the individual strengths of the proposed approach, as well as their interplay with standard near-field subspace methods, we next provide a qualitative comparison, which is complemented by a quantitative evaluation reported in Section~\ref{sec:experiments}. In the following, we focus on the following  measures: {\em Complexity}, {\em Training}, and {\em Favorable Properties}. The comparison is summarized as Table~\ref{tab:discussion comparison}. 

{\bf Complexity:} 
As our algorithms combine \acp{dnn} with subspace methods, their inference complexity is highly coupled with that of standard Near-Field \ac{music}. The latter is dominated by the empirical covariance estimate (complexity of $\mySet{O}(T N^2)$); \ac{evd} (complexity of $\mySet{O}(N^3)$); and evaluation of the \ac{music} spectrum over a grid with $G_{\rm 2D} = G_{\rm \text{A}} \times G_{\rm \text{R}}$ points (complexity of $\mySet{O}(G_{\rm 2D} N^2)$). 
\name~preserves the operation of Near-Field \ac{music}, while replacing the empirical covariance with empirical autocorrelation (complexity of $\mySet{O}(\tau T N^2)$), to which a \ac{cnn} autoencoder is applied. As the number of products of a convolutional layer is proportional to its input size and the number of parameters~\cite{freire2024computational}, its complexity is of an order of $\mySet{O}(C N^2)$, where the coefficient $C$ depends on the number of \ac{dnn} parameters and layers. In the likely case where $N \ll \max(G_{\rm 2D}, C, T)$, the resulting complexity is
\begin{equation}
    \label{eqn:ComplexPMN}
    \mySet{C}_{\rm NFS} = \mySet{O}\left(N^2(\tau T + C + G_{\rm 2D})\right),
\end{equation}
which is often dominated by the grid size $G_{\rm 2D}$.

 \ac{dcdmusic} eliminates the need for exhaustive 2D grid searching by processing the input signal through two autoencoders, followed by the \ac{esprit} algorithm. Unlike \ac{music}, \ac{esprit} solves a \acl{ls} problem and performs an $M$-dimensional \ac{evd} instead of peak searching, at an overall complexity order of $\mySet{O}(N M^2)$. The final step applies 1D-\ac{music} $M$ times over the range dimension alone, using a 1D grid of $G_{\rm \text{R}}$ points. Consequently, the overall complexity order, again assuming $N \ll \max(G_{\rm \text{R}}, C, T)$, is 
\begin{equation}
    \label{eqn:ComplexDCD}
    \mySet{C}_{\rm DCD} = \mySet{O}\left(N^2(\tau T + 2C + M G_{\rm \text{R}})\right).
\end{equation}
Consequently, the complexity difference between \name~in \eqref{eqn:ComplexPMN} and \ac{dcdmusic} in \eqref{eqn:ComplexDCD} depends on the selected grid sizes and the complexity of the autoencoders. 

{\bf Training:} 
Being supervised \ac{ai}-aided algorithms, both \name~and \ac{dcdmusic} require training from labeled data. While training is typically done offline, its complexity is still an important aspect.
For \name, the training procedure is straightforward, and given the novel loss function in \eqref{eq: loss over dataset for musicpeaknet}, it can be trained via standard deep learning methods as in Algorithm~\ref{alg:pmn_training}. The cascaded operation of \ac{dcdmusic} necessitates a more involved three-step training procedure as summarized in Algorithm~\ref{alg:dcd_training}, making it more complex and time-consuming to optimize compared to \name.

{\bf Favorable Properties:} 
Being the main motivation in their formulation, both \name~and \ac{dcdmusic} can successfully apply subspace methods in the presence of challenges \ref{itm:NonCoherent}-\ref{itm:SNR}, where standard subspace methods struggle. These capabilities are numerically demonstrated  in Section~\ref{sec:experiments}. However,  each algorithm achieves these capabilities with a different structure, leading to distinct properties.

For once, \ac{dcdmusic} does not preserve the flow of Near-Field \ac{music}, and separates angle recovery from range estimation. As a result, it cannot handle sources that share the same angle. However, the fact that it learns to compute surrogate covariance matrix under the far-field assumption, makes it well-suited for handling far-field sources and mixed near-far field scenarios (\ref{itm:Mixed}), and can achieve this capability without being trained to do so, as shown in Section~\ref{sec:experiments}. 

While \name~can also handle mixed near-field/far-field sources (\ref{itm:Mixed}), it requires dedicated training using  data comprised with both far and near field sources to do so. However, the fact that it preserves the operation of Near-Field \ac{music} allows it to localize users at similar angles. Moreover, its surrogate "clean" near-field covariance matrix can also be combined with alternative algorithms that require such statistical moments, e.g., generating focused beams via beamforming, as we show in Section~\ref{sec:experiments}.

\begin{table}
\centering
    \renewcommand{\arraystretch}{1.5}
    {\scriptsize
    \begin{tabular}{|p{1.1cm}|p{2.1cm}|p{2.3cm}|p{1.8cm}|}
    \hline
                                  & {\bf Near-Field \ac{music}}                        & {\bf \name}                                                & {\bf \ac{dcdmusic}}   \\ \hline
     Complexity                     & $\mathcal{O}\big(N^2(T + G_{\rm 2D})\big) $ \cellcolor[HTML]{FFEAAD}  & $\mathcal{O}\big(N^2(\tau T + C + G_{\rm 2D})\big)$ \cellcolor[HTML]{FFEAAD}  & $\mathcal{O}\big(N^2(\tau T + 2C + M G_{\rm \text{R}})\big)$ \cellcolor[HTML]{FFEAAD}  \\ \hline
         Training                          
         & Not needed        \cellcolor[HTML]{AAFDB4}                    
         & Standard    \cellcolor[HTML]{FFEAAD}                                     & 3 Stages \cellcolor[HTML]{FF9595}               \\ \hline
    \ref{itm:NonCoherent}-\ref{itm:SNR} 
    & Not suitable  \cellcolor[HTML]{FF9595}                 
    & Suitable             \cellcolor[HTML]{AAFDB4}                                       & Suitable        \cellcolor[HTML]{AAFDB4}          \\ \hline
    Shared angles                       & Suitable    \cellcolor[HTML]{AAFDB4}                    & Suitable \cellcolor[HTML]{AAFDB4}                                                & Not suitable    \cellcolor[HTML]{FF9595}              \\ \hline
    Handle \ref{itm:Mixed}        & Requires adaptation   \cellcolor[HTML]{FFEAAD}        & At additional training      \cellcolor[HTML]{FFEAAD}                                     & Suitable  \cellcolor[HTML]{AAFDB4}                 \\ \hline
    \end{tabular}
}
\caption{Qualitative comparison summary.}
\label{tab:discussion comparison}
\end{table}

{\bf Potential Extensions:}
Our \ac{ai}-aided near-field localization algorithms give rise to multiple avenues for future investigation. While we focus here on localization with \acp{ula}, our methodology can potentially extend to planar arrays, enabling 3D near-field localization~\cite{chen2025near}, as well as to sparse arrays~\cite{vaidyanathan2010sparse}, leveraging data to cope with the inherent challenge in calibrating virtual array elements. Moreover, the fact that we cast near-field localization as a machine learning model facilitates its joint learning along with likely downstream processing, such as tracking~\cite{guerra2021near}. This system-level perspective be potentially used to learn to provide features most useful for tracking, leverage tracking confidence for online adaptation, and even jointly learning to localize and track~\cite{buchnik2023latent}. These extensions are all left for future study.


\section{Experimental Study}
\label{sec:experiments}
This section  numerical evaluates our proposed \ac{ai}-aided near-field localization algorithms. We first detail the experimental setup  in Subsection~\ref{ssec:ExpSetup}. Then, we report our numerical results, evaluating estimation accuracy and the ability to generate interpretable spectra in Subsections~\ref{ssec:ExpAcc}-\ref{ssec:ExpInterp}, respectively. 

\begin{figure*}
    \centering
    \begin{subfigure}{0.32\textwidth}
        \centering
        \includegraphics[width=\textwidth]{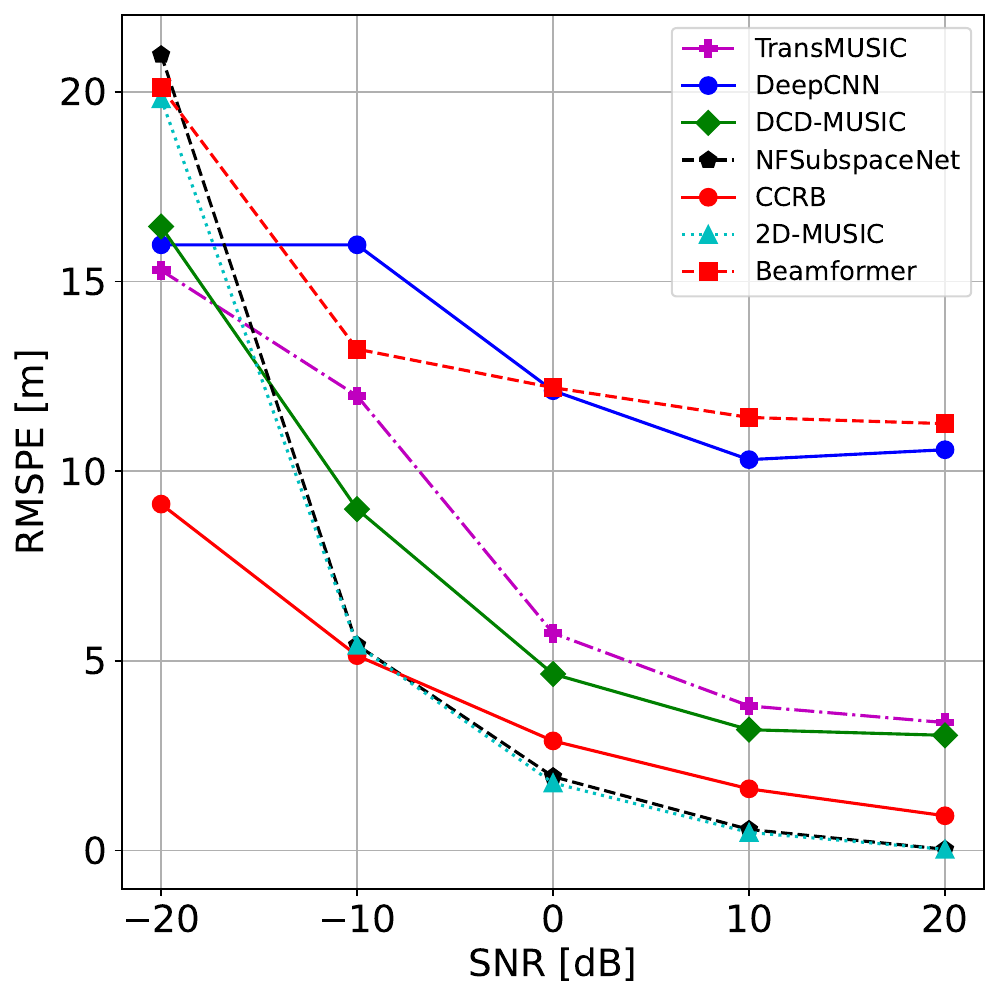}
        \caption{Non-coherent sources.}
        \label{fig:2 non-coherent sources}
    \end{subfigure}%
    \begin{subfigure}{0.32\textwidth}
        \centering
        \includegraphics[width=\textwidth]{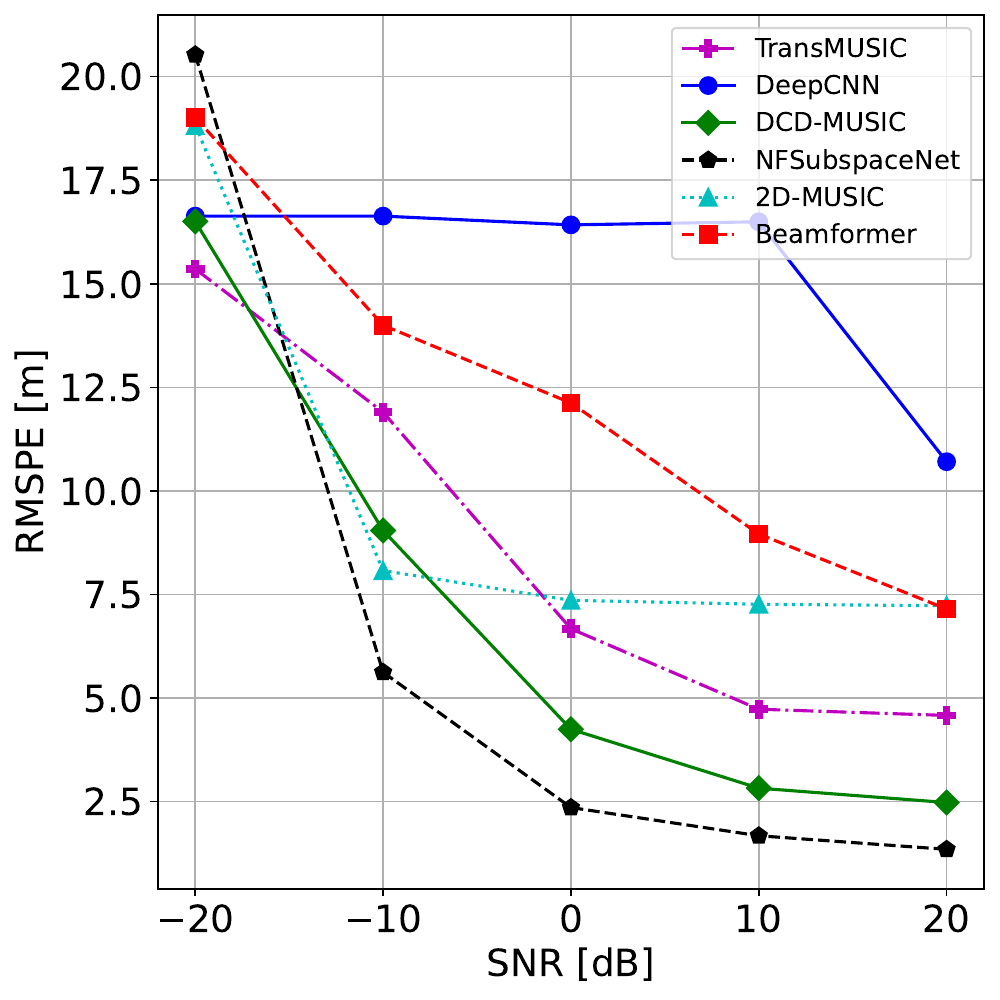}
        \caption{Coherent sources.}
        \label{fig:2 coherent sources}
    \end{subfigure}
    \begin{subfigure}{0.32\textwidth}
        \centering
        \includegraphics[width=\textwidth]{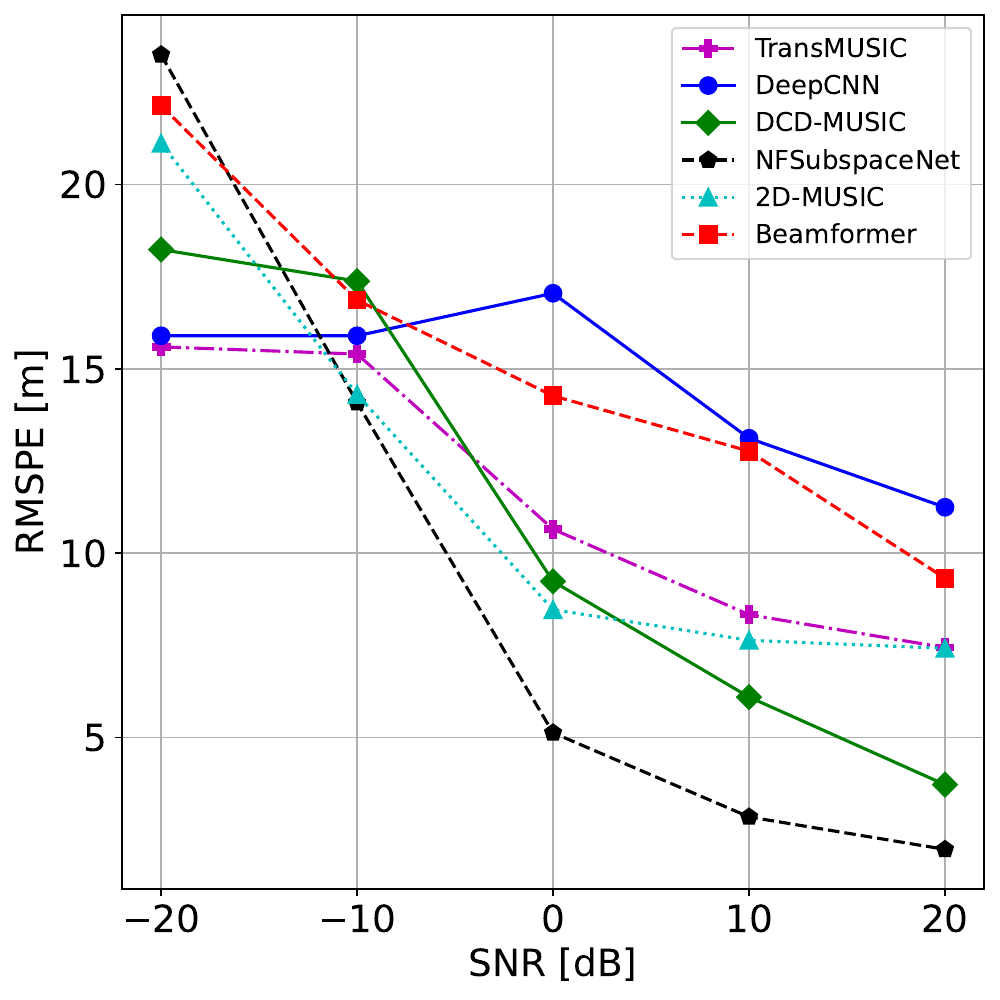}
        \caption{Coherent sources \& $T=10$.}
        \label{fig:2 coherent sources with low number of snapshots}
    \end{subfigure}
    \vspace{0.4cm}
    \caption{\ac{rmspe} vs. \acs{snr}, $M=2$ sources.}
    \label{fig: 2 sources results}
\end{figure*}

\subsection{Experimental Setup}
\label{ssec:ExpSetup}
\subsubsection{Signal Model} 
We simulate two different \ac{ula} configurations: 
$(i)$ The setting used in Subsections~\ref{ssec:ExpAcc}-\ref{ssec:ExpInterp} employs an array with $N = 15$ sensors, spaced at half the wavelength at a carrier frequency of $300$ MHz;
$(ii)$ The large-scale setting evaluated in Subsection~\ref{ssec:ExpLarge} uses an array with $N=64$ elements with half wavelength spacing at $5$ GHz. 
In all settings, the sources' \acp{doa} and ranges are uniformly drawn from $\big[\frac{-\pi}{3}, \frac{\pi}{3}\big]$ and from the Fresnel limit up to half the Fraunhofer limit, respectively. Unless stated otherwise, the trajectory length is set to $T = 100$, the array is assumed to be fully calibrated, and both the signal and noise are modeled as complex normal distributions with zero mean and unit covariance. For the coherent case, the sources transmit the same signal, creating fully correlated sources.

\subsubsection{Evaluation Metrics}
We evaluate the  \ac{rmspe} of the position error, which is therefore reported in meters, under varying key parameters, including the \ac{snr}, the number of snapshots and the number of sources. Further, we assess  robustness in the presence of coherent sources, and array miscalibration, where the exact locations of the sensor elements are unknown to the tested algorithms. 
In addition to \ac{rmspe}, we also assess  the algorithms' ability to estimate the number of sources using the different suggested techniques. 
The interpretability of our solutions is analyzed through MUSIC spectrum and the beampattern.

\begin{table*}
\centering
\adjustbox{max width=\textwidth}{
\begin{tabular}{|c|c|c|c|c|c|c|c|}
\hline
\multirow{2}{*}{\textbf{Scenario}} & \multirow{2}{*}{\textbf{Method}} & \multicolumn{2}{c|}{\textbf{2D-\ac{music}}} & \multicolumn{2}{c|}{\textbf{\name}} & \multicolumn{2}{c|}{\textbf{\ac{dcdmusic}}} \\ \cline{3-8}
 & & Acc.[\%]& RMSPE[m]& Acc.[\%]& RMSPE[m]& Acc.[\%]& RMSPE[m]\\
\hline\hline
\multirow{4}{*}{Non Coherent} 
& None      & -            & \multirow{4}{*}{\bgreen{0.674}} & -           & 0.716  & -             & 4.578  \\ \cline{2-3} \cline{5-8}
& MDL       & \bgreen{100} &                                 & 99.5        & 0.712  & 88.6          & 4.654 \\ \cline{2-3} \cline{5-8}
& AIC       & 92.2         &                                 & \bgreen{94} & 0.723  & 92.5          & 5.349 \\ \cline{2-3} \cline{5-8}
& Threshold & 32.7         &                                 & 34.7        & 0.72   & \bgreen{40.2} & 5.5 \\ 
\hline\hline
\multirow{4}{*}{Coherent} 
& None      & - & \multirow{4}{*}{13.758} & -             & \bgreen{7.665} & -     & 10.463 \\ \cline{2-3} \cline{5-8}
& MDL       & 0 &                         & 32            & \bgreen{7.832} & 12.1  & 10.231 \\ \cline{2-3} \cline{5-8}
& AIC       & 0 &                         & \bgreen{58.5} & \bgreen{7.303} & 48.5  & 10.231 \\ \cline{2-3} \cline{5-8}
& Threshold & 0 &                         & 34.2          & \bgreen{7.688} & 27.2  & 10.622 \\
\hline
\end{tabular}
}
\caption{Performance comparison of different model order estimation methods and regularization techniques.}
\label{tab:performance_comparison_v2}
\end{table*}

\subsubsection{Localization Algorithms}
We implement \name~and \ac{dcdmusic} using a \ac{cnn} autoencoder architecture with $3$ encoding and decoding convolutional layers. 
%
We compare our results to different model-based algorithms, including  Near-Field \ac{music} (termed {\em 2D \ac{music}}), which employs \ac{sps} for the coherent case \cite{wang1994spatial}; and standard beamforming~ \cite{stoica2005spectral}. For non-coherent cases, we also report the \ac{ccrb}, using its derivation for non-coherent sources in  \cite{el2010conditionalcrb}.
As for data-driven benchmarks, we use the \ac{transmusic} architecture~\cite{ji2024transmusic} and the \ac{cnn} based model suggested~\cite{papageorgiou2021deepcnn} (which we coin {\em DeepCNN}), adapting them to be suitable for the near-field scenario. 
In settings with unknown number of sources, we used \ac{aic}, \ac{mdl}, or  thresholding to estimate $M$.

All methods involving grid searching use the same resolution of $0.5^\circ$  in $\big[-60^\circ, 60^\circ\big]$ and $0.5$ meters between the Fresnel and half the Fraunhofer limits.
Unless stated otherwise, the training dataset comprises $|\mySet{D}| = 4096$ samples, we focus on a limited training dataset to match a real life system which labeld data is not easily obtained, and the algorithms were evaluated on a test dataset with 410 different source locations over the specified possible values above. The full details of our architecture and training hyperparameters can be found in our GitHub repository at \url{https://github.com/ShlezingerLab/AI-Subspace-Methods}.

\subsection{Estimation Accuracy}
\label{ssec:ExpAcc}
\subsubsection{Fixed Number of Sources}
Our first evaluation compares the considered localization algorithms in the case where the number of sources is fixed at $M = 2$. The \ac{rmspe} results are presented in Fig.~\ref{fig: 2 sources results} for non-coherent and coherent sources, as well as for a scenario with coherent sources and a short trajectory of only $T = 10$ samples.
Fig.~\ref{fig: 2 sources results} indicates that when conditions \ref{itm:NonCoherent} and \ref{itm:SNR} do not hold,  2D-\ac{music} performs well, similar to data-driven approaches. However, when dealing with coherent sources  or an insufficient number of snapshots, the model-based algorithm fails to localize accurately. In contrast, both \name~ and \ac{dcdmusic} demonstrate robust performance under these challenging conditions.
{Additionally, due to their high parameterization, both \ac{transmusic} and Deep\ac{cnn}, struggle to generalize when trained with only $4096$ samples. Specifically,  Deep\ac{cnn} shows superior performances only for low \ac{snr} values. }

\subsubsection{Mixed Number of Sources}
We proceed to a more difficult setting in which the number of sources $M$ varies between 2 and 8 sources, drawn uniformly for each sample in the training and test datasets. 
Additionally, we evaluated different regularization methods to determine which provides better localization and model order estimation for such scenarios. As a baseline, we also included a model-based method for model-order estimation. The results are presented in Table \ref{tab:performance_comparison_v2}. 
{A key observation here is that  using the model order selection test directly on the surrogate covariance yields similar results to the classic use of empirical covariance when the sources are non-coherent. For the coherent case, we can achieve up to $60~\%$ accuracy, where the model-based approaches fail, while also achieving improved \ac{rmspe} performance.}
Note that the results of \ac{transmusic}, which uses an additional \ac{dnn} for estimating $M$, were not included in the table  due to observing a major reduction with localization performance when training the source estimation branch as suggests in \cite{ji2024transmusic}. 

Moreover,  the regularization introduced to facilitate model-order selection does not necessarily degrade localization accuracy, particularly in more challenging settings with coherent sources, possibly owing to its further encouraging the surrogate covariance to be suitable to downstream subspace-based processing. {Specifically, when considering the coherent case with \ac{aic}, \name~manages to improve its localization comparing to training without any regularization.}

\begin{figure}
    \begin{subfigure}{0.24\textwidth}
        \includegraphics[width=\textwidth]{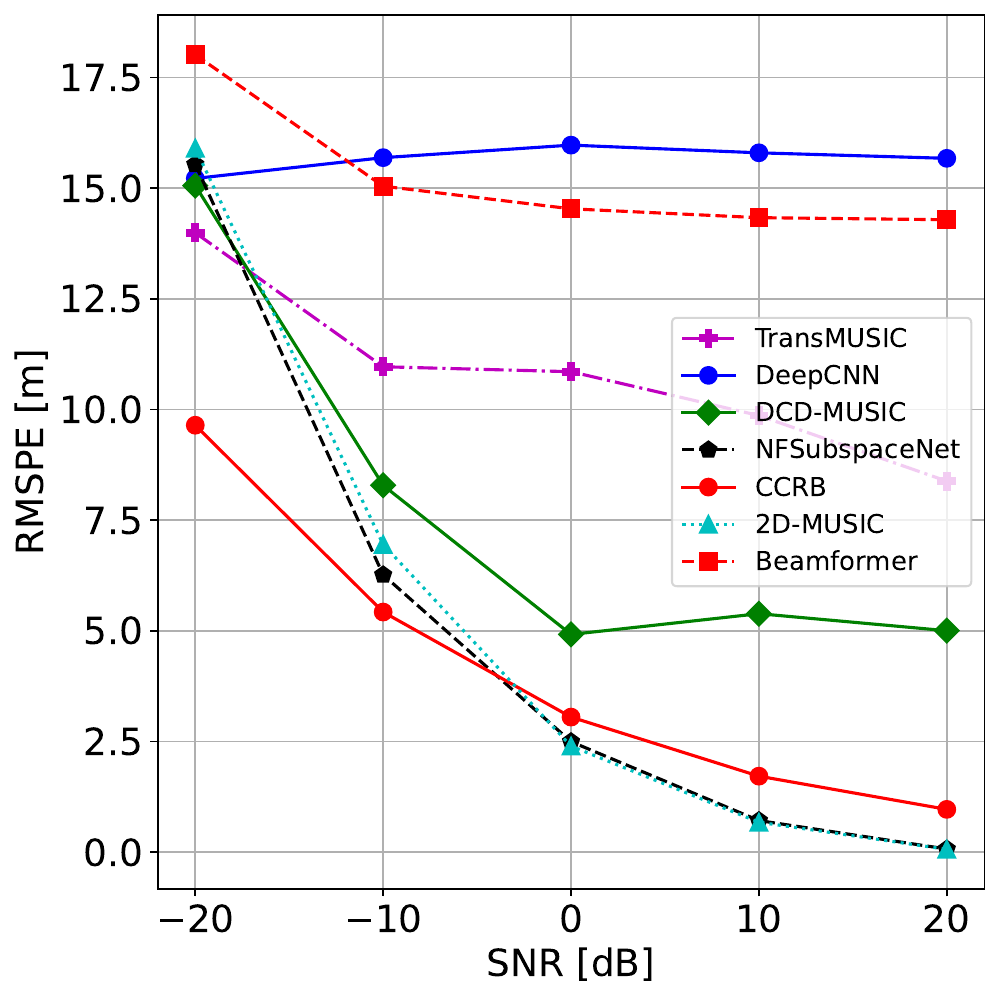}
        \caption{\ac{rmspe} VS. SNR.}
        \label{fig:2_8 non-coherent sources rmspe}
    \end{subfigure}%
    \hspace{0.05cm}
    \begin{subfigure}{0.24\textwidth}
        \includegraphics[width=\textwidth]{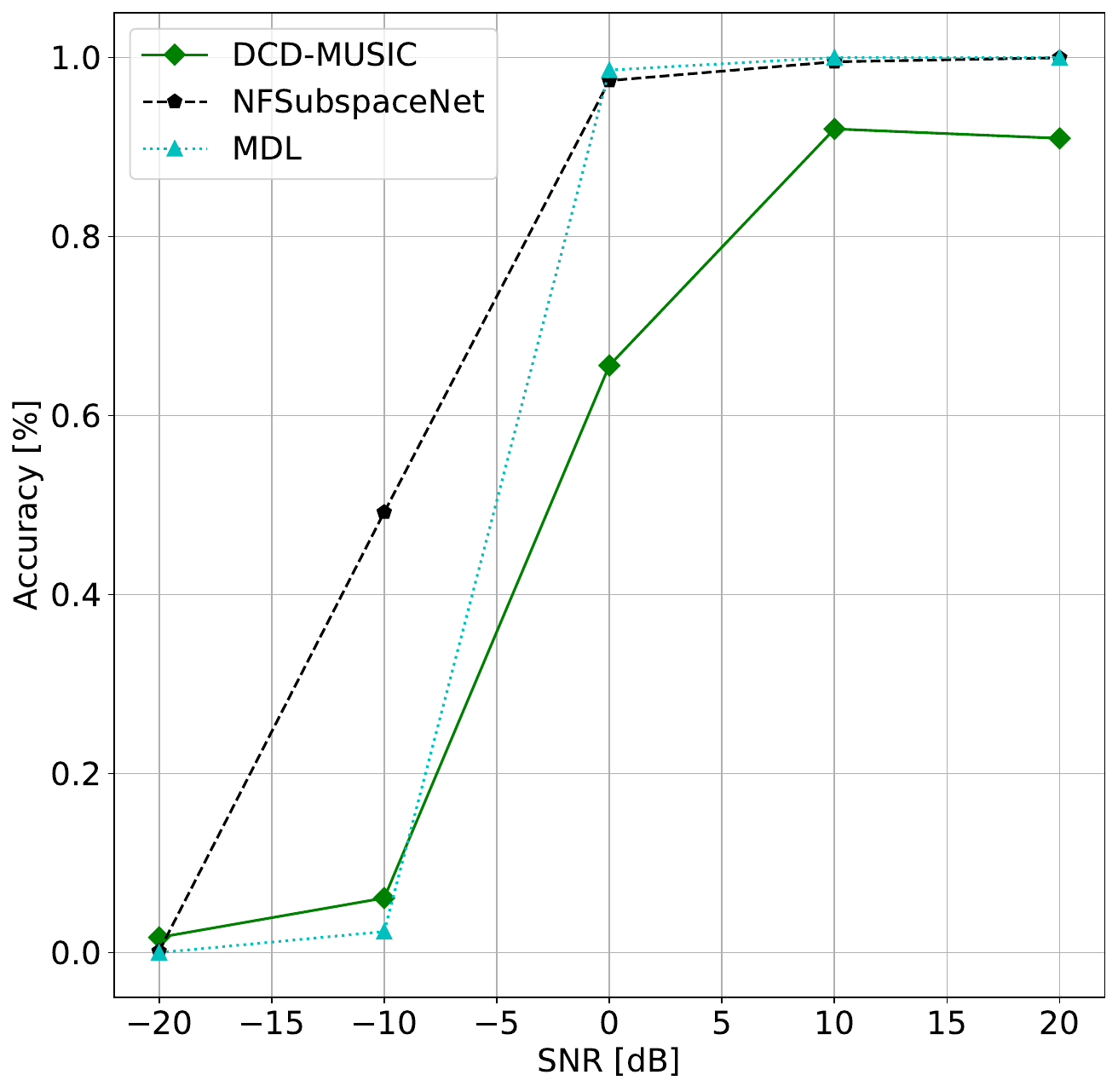}
        \caption{ACC. VS. SNR.}
        \label{fig:2_8 non-coherent sources acc}
    \end{subfigure}
    \vspace{0.1cm}
    \caption{Mixed number of non-coherent sources.}
    \label{fig: 2_8 non-coherent sources results}
\end{figure}

\begin{figure}
    \begin{subfigure}{0.24\textwidth}
        \includegraphics[width=\textwidth]{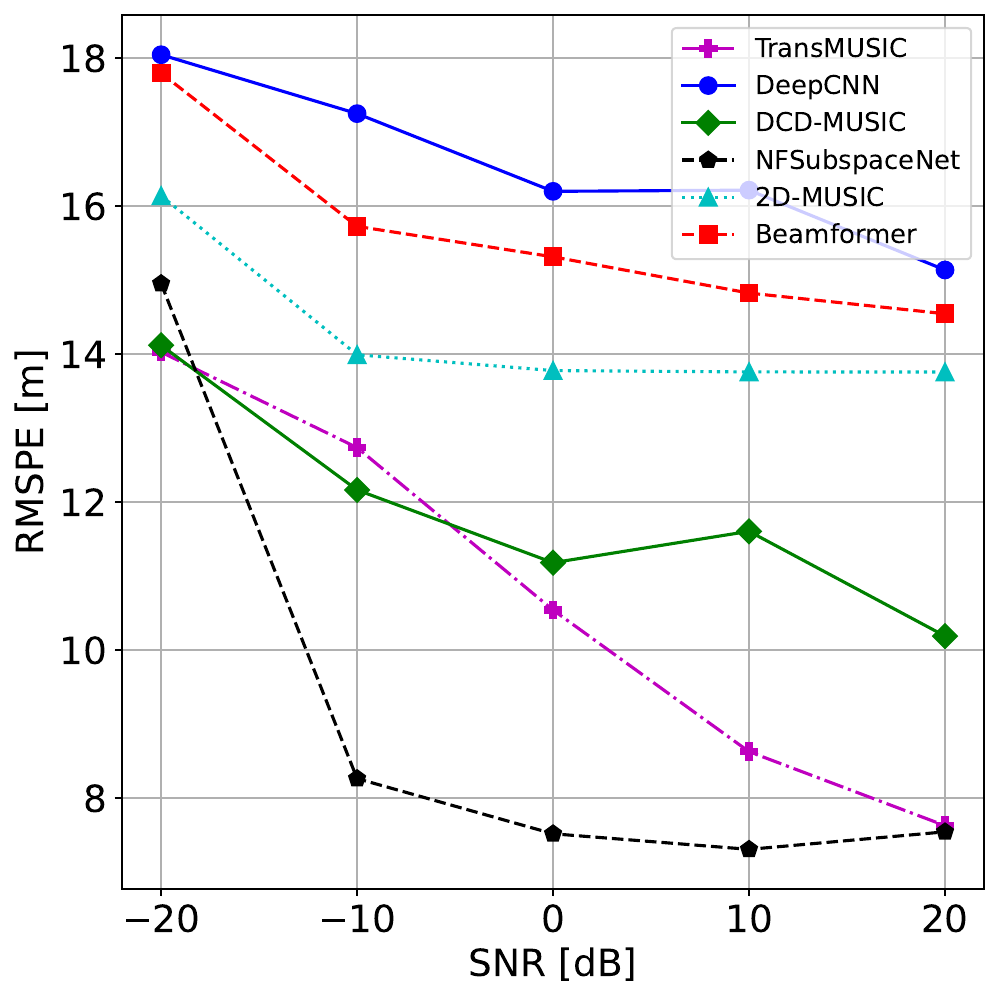}
        \caption{\ac{rmspe} VS. SNR.}
        \label{fig:2_8 coherent sources rmspe}
    \end{subfigure}%
    \hspace{0.05cm}
    \begin{subfigure}{0.24\textwidth}
        \includegraphics[width=\textwidth]{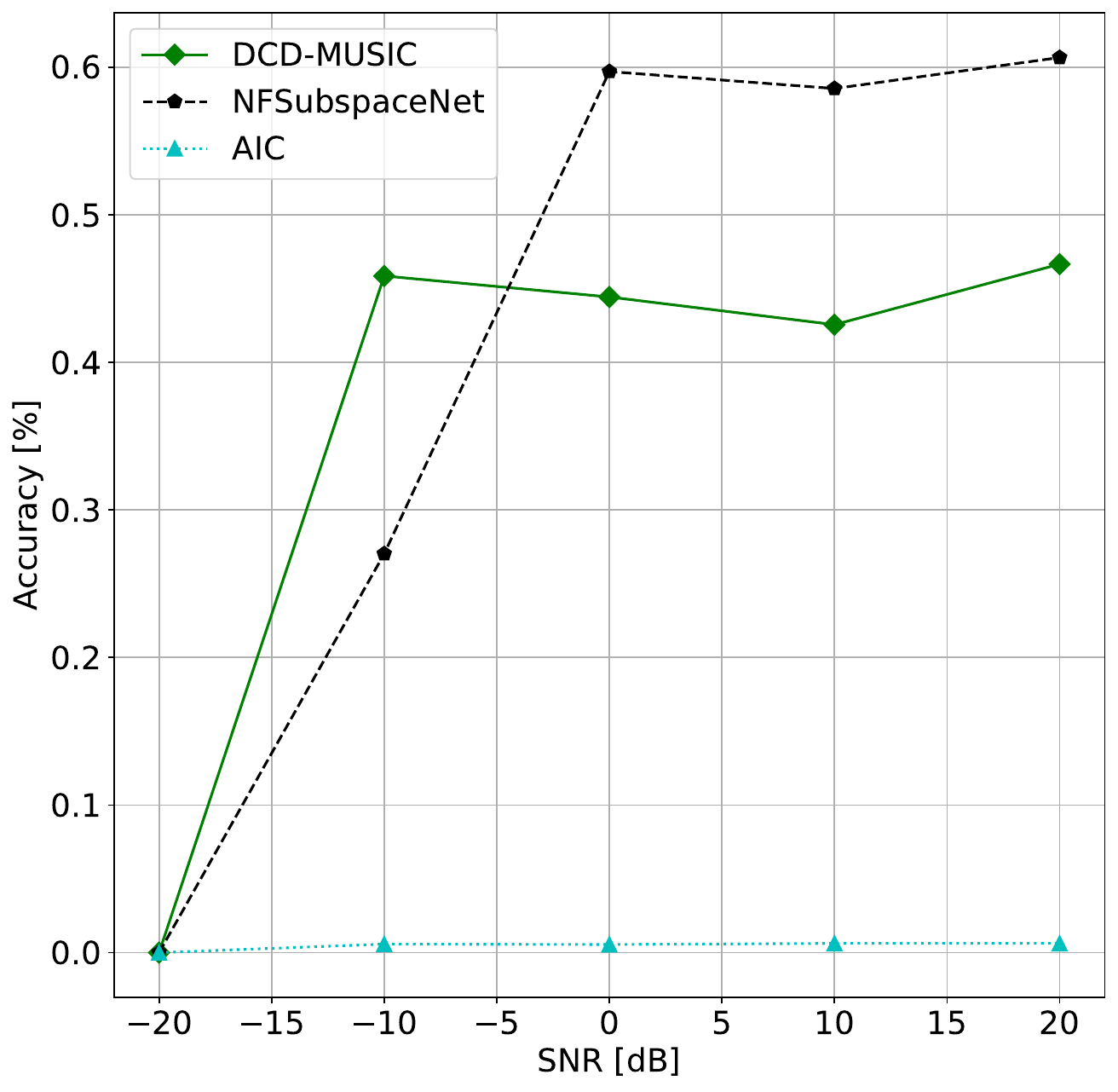}
        \caption{ACC. VS. SNR.}
        \label{fig:2_8 coherent sources acc}
    \end{subfigure}
    \vspace{0.1cm}
    \caption{Mixed number of coherent sources.}
    \label{fig: 2_8 coherent sources results}
\end{figure}

The ability of our algorithms to simultaneously localize and identify the number of sources is further evident when evaluating these metrics in both non-coherent and fully coherent scenarios for different \ac{snr} values. In this scenario, the data size is $|\mySet{D}| = 40,000$ {and the test dataset includes $4,000$ samples}. The results are reported in Fig.~\ref{fig: 2_8 non-coherent sources results} and Fig.~\ref{fig: 2_8 coherent sources results}. Based on the findings in Table~\ref{tab:performance_comparison_v2}, {we use \ac{mdl} and \ac{aic}, for the non-coherent and coherent cases respectively for \name, and \ac{aic} for \ac{dcdmusic}, as the model order estimation method as well as the regularization term in the training stage.
The results suggest that for non-coherent case, both data-driven and model-based methods successfully localize up to 8 sources, with the best performance in terms of both \ac{rmspe} and localization accuracy achieved using using \name. For the coherent case, the data-driven algorithms outperform the model-based ones, and the lowest \ac{rmspe} achieved by the \name~at $\text{SNR} = 20 \text{dB}$ but with only a minor  gap from \ac{transmusic}, that was able to localized better than \ac{dcdmusic}. \name~and \ac{dcdmusic} were able to partly estimate the number of sources with up to 60\% accuracy.}

\subsubsection{Miscalibarion and Limited Snapshots}
\begin{figure}
    \begin{subfigure}{0.24\textwidth}
        \includegraphics[width=\textwidth]{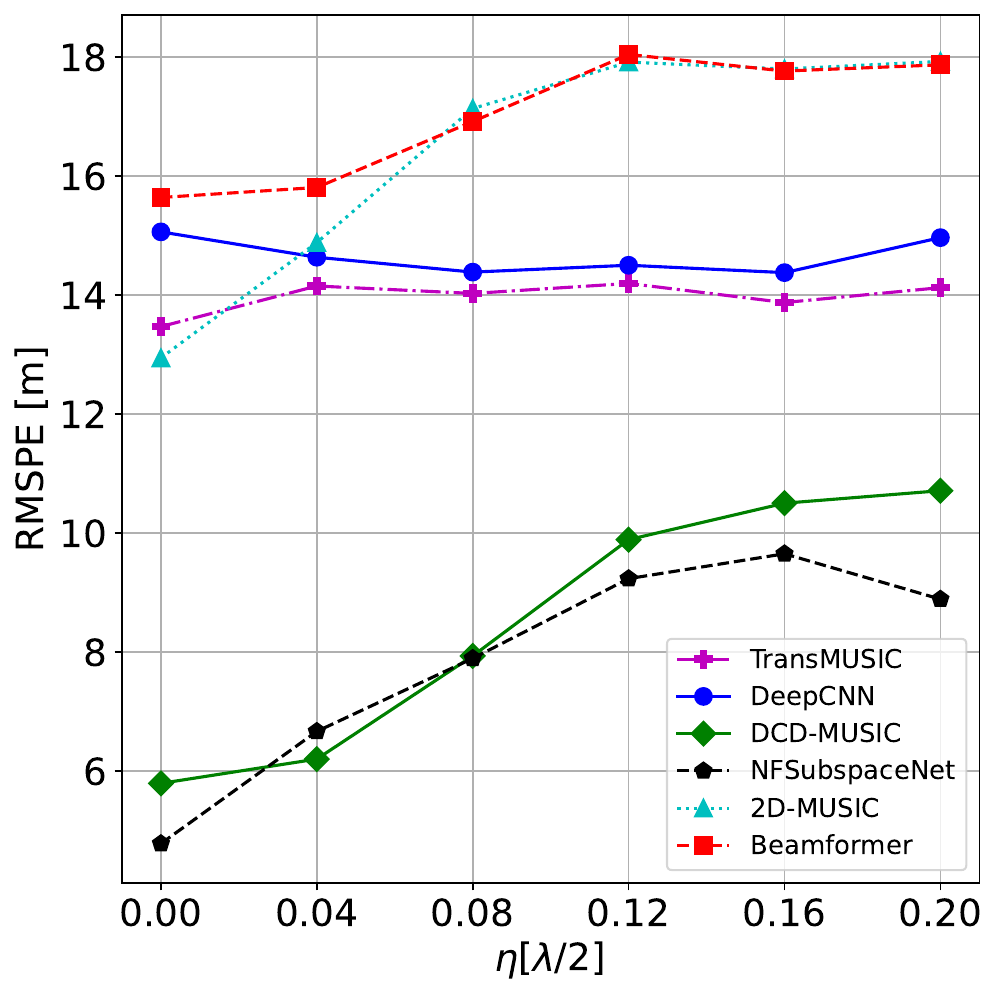}
        \caption{\ac{rmspe} vs. $\eta$.}
        \label{fig:3 coherent sources miscalibrated}
    \end{subfigure}%
    \hspace{0.05cm}
    \begin{subfigure}{0.24\textwidth}
        \includegraphics[width=\textwidth]{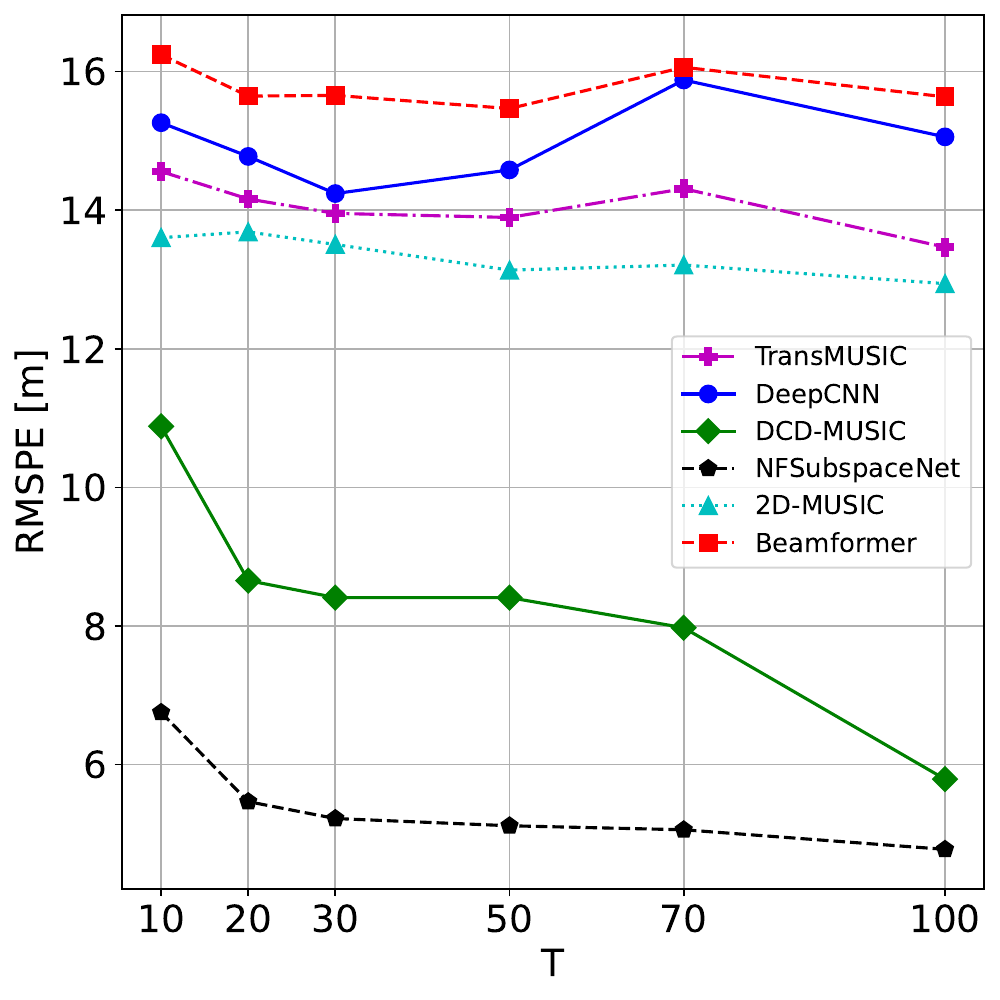}
        \caption{\ac{rmspe}. vs. $T$.}
        \label{fig:3 coherent sources snapshots}
    \end{subfigure}
    \vspace{0.1cm}
    \caption{\ac{rmspe}  vs. $T$ and $\eta$, 3 coherent sources.}
    \label{fig: 3 coherent sources miscal and snap}
\end{figure}

We next consider challenging settings induced by non-calibrated arrays (\ref{itm:MisCalib}) and few snapshots (\ref{itm:SNR}). The former is simulated by adding  noise distributed uniformly in $[-\eta, \eta]$ to each sensor location.
The simulation results for these scenarios with $M=3$ sources at 0 dB \ac{snr} are reported in  Fig.~\ref{fig: 3 coherent sources miscal and snap}. Observing Fig.~\ref{fig: 3 coherent sources miscal and snap}, we note that  our algorithms successfully cope  with miscalibrated array, with both \name~and \ac{dcdmusic} consistently achieving the most accurate localization. We can also see that the ability of \name~to overcome insufficient number of snapshot, only $T=10$.

\begin{table}

\footnotesize
\centering
\renewcommand{\arraystretch}{1.5} 
\adjustbox{max width=\columnwidth}{
\begin{tabular}{|c|c|c|c|} \hline 
    \textbf{Sources Range} & \textbf{2D-\ac{music}} & \textbf{\name} & \textbf{\ac{dcdmusic}} \\ \hline 
    $\text{Fresnel} : 0.5 \cdot \text{Fraunhofer}$ & ${\bf 4.527\cdot 10^{-5}}$ & $5.281\cdot 10^{-5}$ & $1.347\cdot 10^{-2 }$
    \\ (Fresnel approx.) & & & \\ \hline 
    $\text{Fresnel} : 0.5 \cdot \text{Fraunhofer}$ & $5.395\cdot 10^{-2}$ & ${4.154\cdot 10^{-2}}$ & ${\bf 1.66\cdot 10^{-2}}$ \\ \hline 
    $\text{Fresnel} : 1   \cdot \text{Fraunhofer}$ & $1.124\cdot 10^{-1}$ & $1.739\cdot 10^{-1}$ & ${\bf 1.174\cdot 10^{-2}}$ \\ \hline 
    $\text{Fresnel} : 2   \cdot \text{Fraunhofer}$ & $2.485\cdot 10^{-1}$ & $2.75\cdot 10^{-1}$ & ${\bf 1.045\cdot 10^{-2}}$ \\ \hline 
    $\text{Fresnel} : 5   \cdot \text{Fraunhofer}$ & $3.709\cdot 10^{-1}$ & $3.811\cdot 10^{-1}$ & ${\bf 9.583\cdot 10^{-3}}$ \\ \hline 
    $\text{Fresnel} : 10   \cdot \text{Fraunhofer}$ & $3.753\cdot 10^{-1}$ & $3.708\cdot 10^{-1}$ &  ${\bf 9.337\cdot 10^{-3}}$ \\ \hline
\end{tabular}
}
\caption{Angle RMSPE in radians for different ranges.}
\label{tab:mix far and near field sources RMSE over the angle.}
\vspace{-0.1cm}
\end{table}

\begin{figure*}
    \centering
    \begin{subfigure}{0.32\textwidth}
        \centering
        \includegraphics[width=\textwidth]{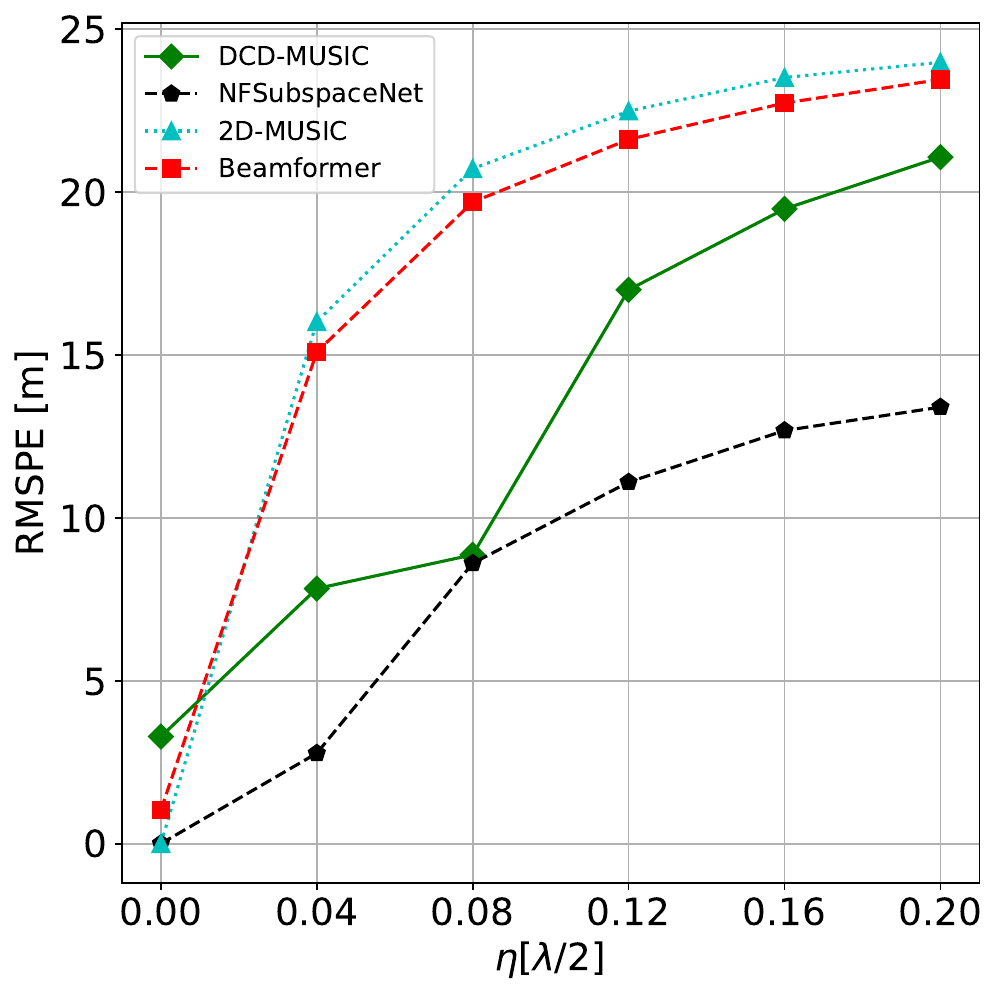}
        \caption{Postion \ac{rmspe}}
        \label{fig:large_ula_position}
    \end{subfigure}%
    \begin{subfigure}{0.32\textwidth}
        \centering
        \includegraphics[width=\textwidth]{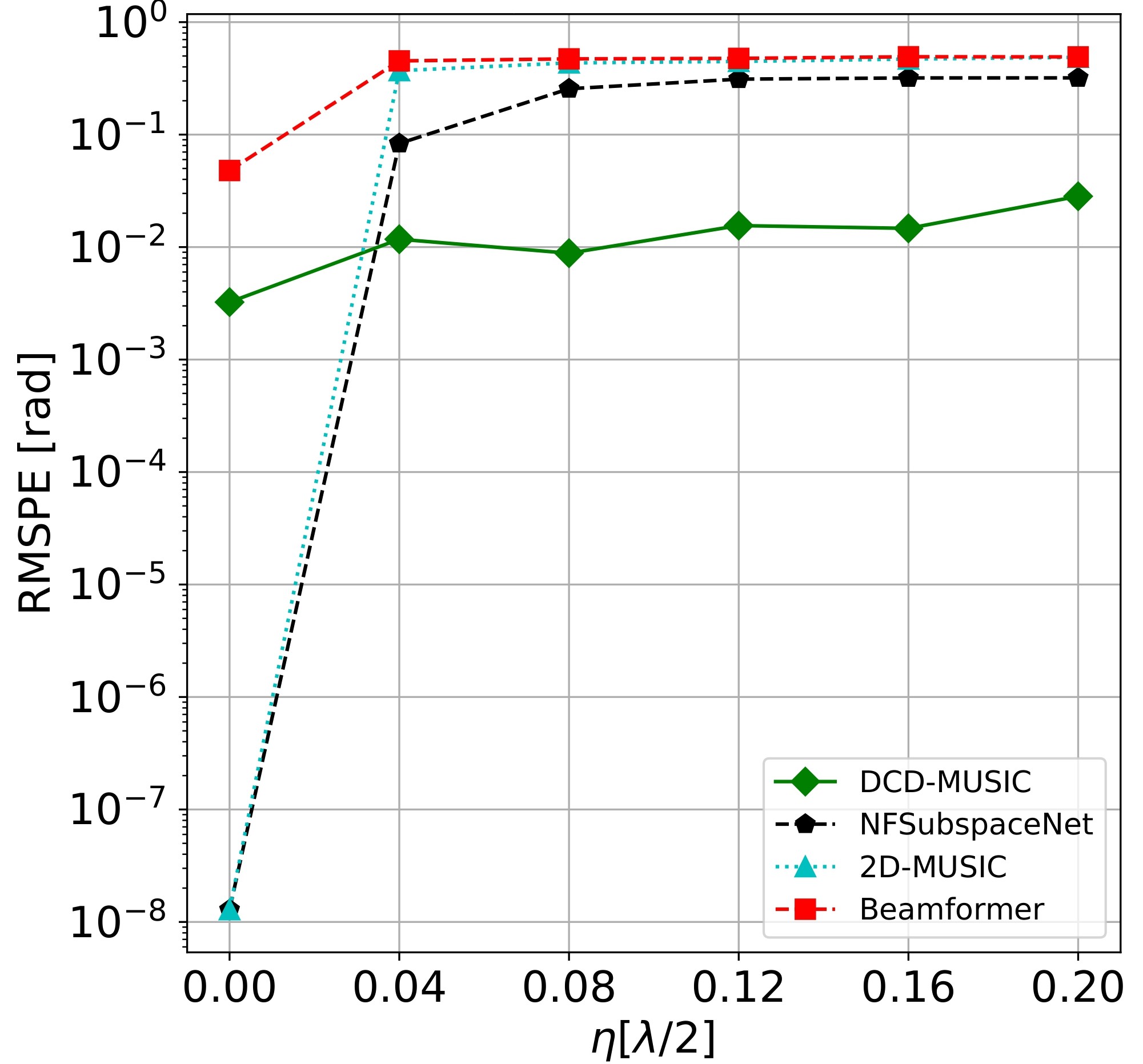}
        \caption{Angle \ac{rmspe}}
        \label{fig:large_ula_angle}
    \end{subfigure}
    \begin{subfigure}{0.32\textwidth}
        \centering
        \includegraphics[width=\textwidth]{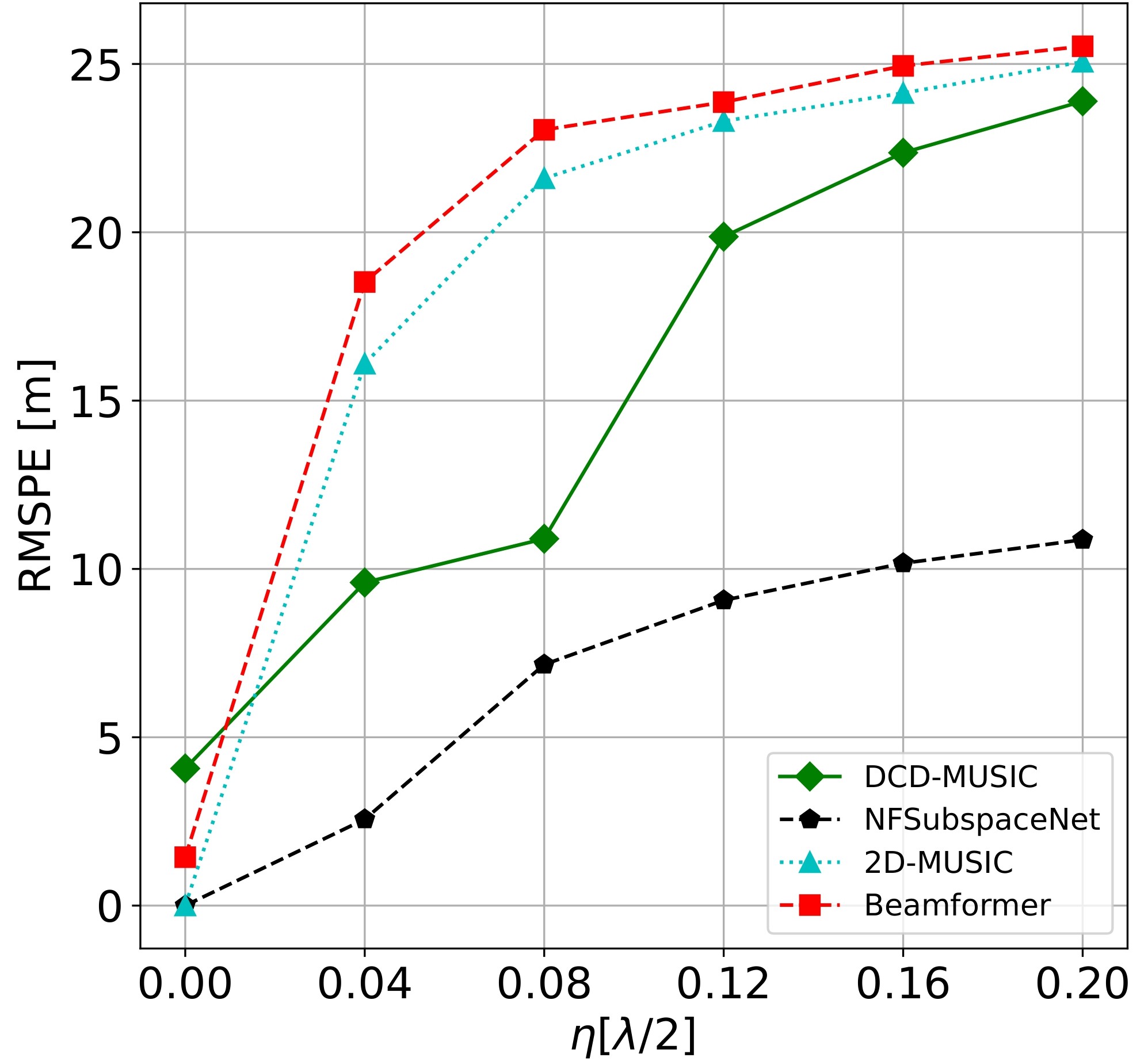}
        \caption{Range \ac{rmspe}}
        \label{fig:large_ula_range}
    \end{subfigure}
    \vspace{0.4cm}
    \caption{\ac{rmspe} vs. Calibration error, $M=2$ non-coherent sources.}
    \label{fig:large_ula_v2}
\end{figure*}

\subsubsection{Mixed Near-Field/Far-Field Sources}
We conclude the section by considering a mixture of sources located in both near-field and far-field regions (\ref{itm:Mixed}). To simulate this scenario, we used the full steering vector model of a \ac{ula}, without relying on Fraunhofer or Fresnel approximations, given as \cite{Guanghui2020, zuo2020subspace} 
\begin{equation*}
    [\myVec{a}(\theta_{m}, \rho_{m})]_n = e^{-j\cdot \frac{2\pi}{\lambda} \rho_{m} \left(1 - \sqrt{1-2\cdot \frac{\lambda\cdot n}{2 \cdot \rho_{m}} \sin{\theta_{m}} + (\frac{\lambda\cdot n}{2 \cdot \rho_{m}}})^{2} \right)}.
\end{equation*}
We compare \ac{dcdmusic} and \name~with 2D-\ac{music}. All methods employ near-field approximations as in \eqref{eq: steering vec model}.
The  scenario consists of $M = 2$ non-coherent sources with an \ac{snr} of $20$ dB. The \ac{ai}-aided models were trained on data simulated using the Fresnel approximation rather than the full steering vector model. Additionally, the search grid for the range was constrained to values within the Fraunhofer limit.

The results for this setting are reported in Table \ref{tab:mix far and near field sources RMSE over the angle.}. The first two rows illustrate the differences arising from using different steering vector models for data generation. As expected, all methods exhibit some degradation in performance. When examining additional sources beyond the Fraunhofer limit, we observe that \ac{dcdmusic} actually improves, owing to its ability to map near-field sources into a far-field surrogate covariance matrix.
In contrast, both \name~and 2D-\ac{music} struggle in these cases. The application of the Fresnel approximation to both near-field and far-field sources in \ac{music} introduces errors due to the Taylor series residual, while \name, having been trained solely on near-field sources with the Fresnel approximation, encounters difficulties when generalizing to far-field conditions. Remarkably, \ac{dcdmusic} demonstrates strong performance in zero-shot inference for mixed far-field and near-field sources.

\subsubsection{Large-Scale \ac{ula}}
\label{ssec:ExpLarge} 
A key motivation for developing near-field localization algorithms arises due to the anticipated deployment of large antenna arrays \cite{wang2024tutorial}. To demonstrate the applicability of our method, we simulated a large-scale \ac{ula} with $N = 64$ elements and a carrier frequency of 5 GHz. In this scenario we tested the algorithms' ability  to cope with calibration error (modeled as a randomized shifts in each sensor location) with \ac{snr} of 20 dB. Here, the Fresnel and Fraunhofer distances, calculated using the array diameter as per \eqref{eq: fresnel region}, result in a Fresnel distance of about 3 meters and Fraunhofer limit of about 120 meters. As this changes the range values, we use a larger dataset of size $|\mySet{D}| = 10^5$ for training. 
The data-driven benchmarks failed to learn in such large scale settings from such limited data, and are thus not reported here.

The results for this scenario are presented in Fig.~\ref{fig:large_ula_v2}. Here, in addition to position results, we also report the individual accuracy in recovering the angle and range,  to examine  the differences in methods performances.
In general, we can see that \ac{dcdmusic} is able to easily adapt to mis-calibrated array, achieving consistent performances for different strength of calibration error. As expected,  both \name~and \ac{music} achieve better results when the array is fully calibrated ($\eta = 0$). In addition, we can see that \name~is most accurate in recovering the position, which likely follows from  its capability to correctly estimate the range. 

\begin{figure*}
    \centering
    \begin{minipage}{0.70\textwidth} 
        \centering
        \begin{subfigure}[t]{0.49\textwidth}
            \centering
            \includegraphics[width=\linewidth]{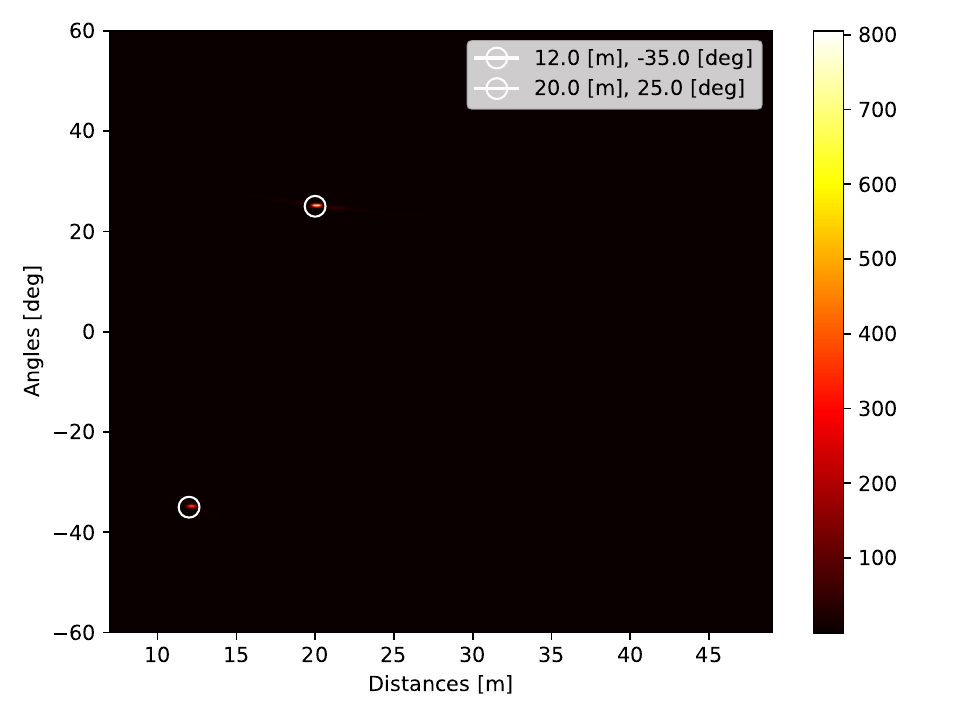}
            \caption{2D-\ac{music} spectrum}
        \end{subfigure}
        \hfill
        \begin{subfigure}[t]{0.49\textwidth}
            \centering
            \includegraphics[width=\linewidth]{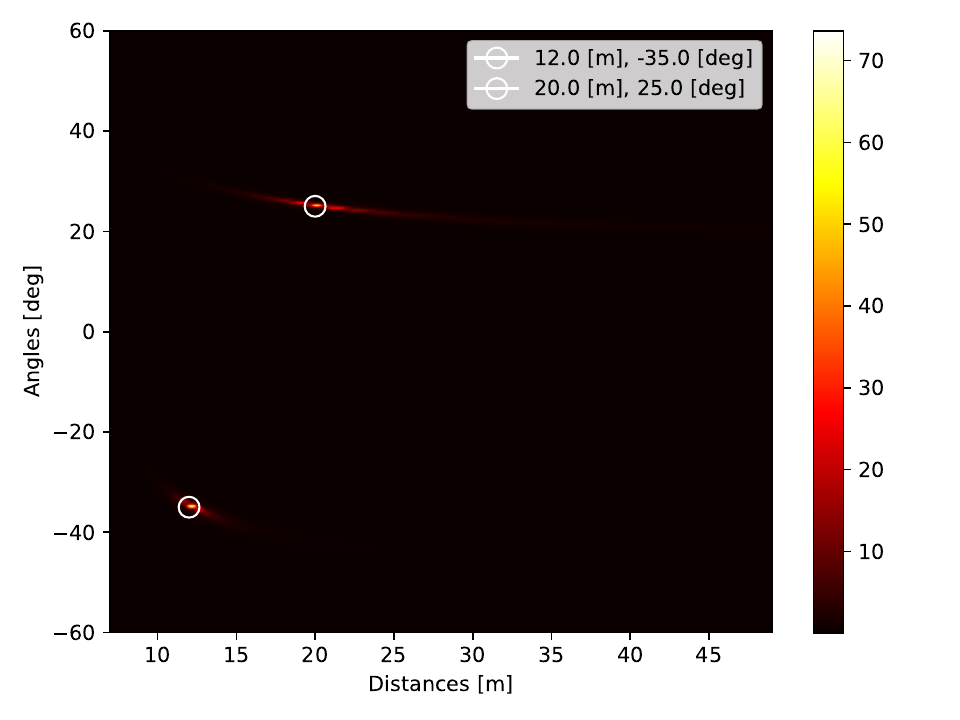}
            \caption{\name ~\ac{music} spectrum}
        \end{subfigure}
    \end{minipage}
    \hfill
    \begin{minipage}{0.25\textwidth} 
        \centering
        \begin{subfigure}[t]{\linewidth}
            \centering
            \includegraphics[width=\linewidth, height=3cm]{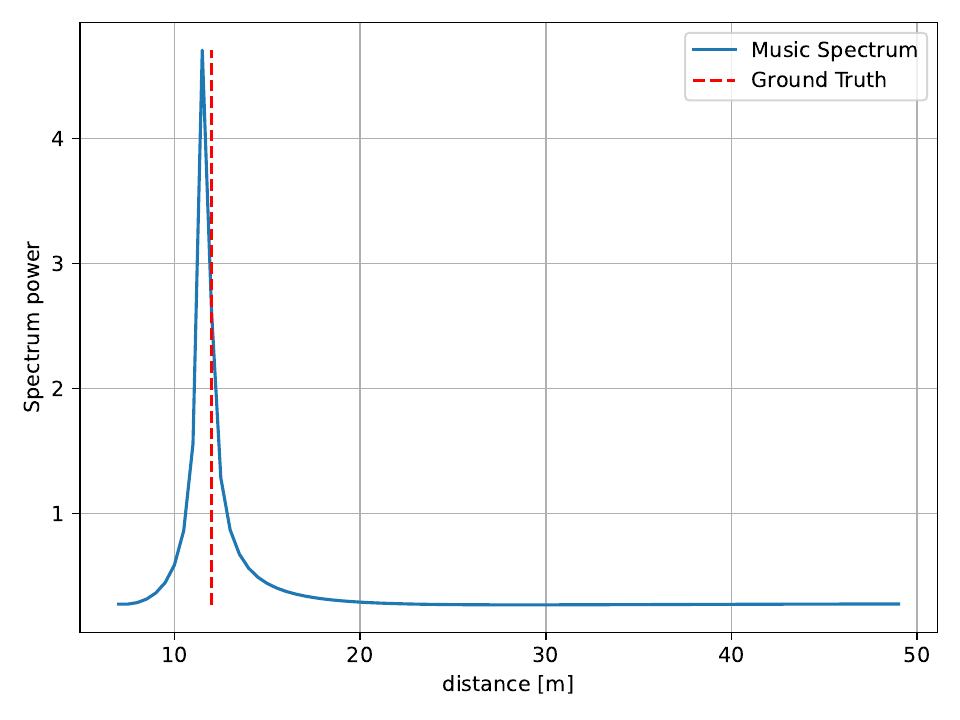}
            \caption{\ac{dcdmusic} spec at $\theta = -35^\circ$}
        \end{subfigure}
        
        \vspace{0.3cm} 
        
        \begin{subfigure}[t]{\linewidth}
            \centering
            \includegraphics[width=\linewidth, height=3cm]{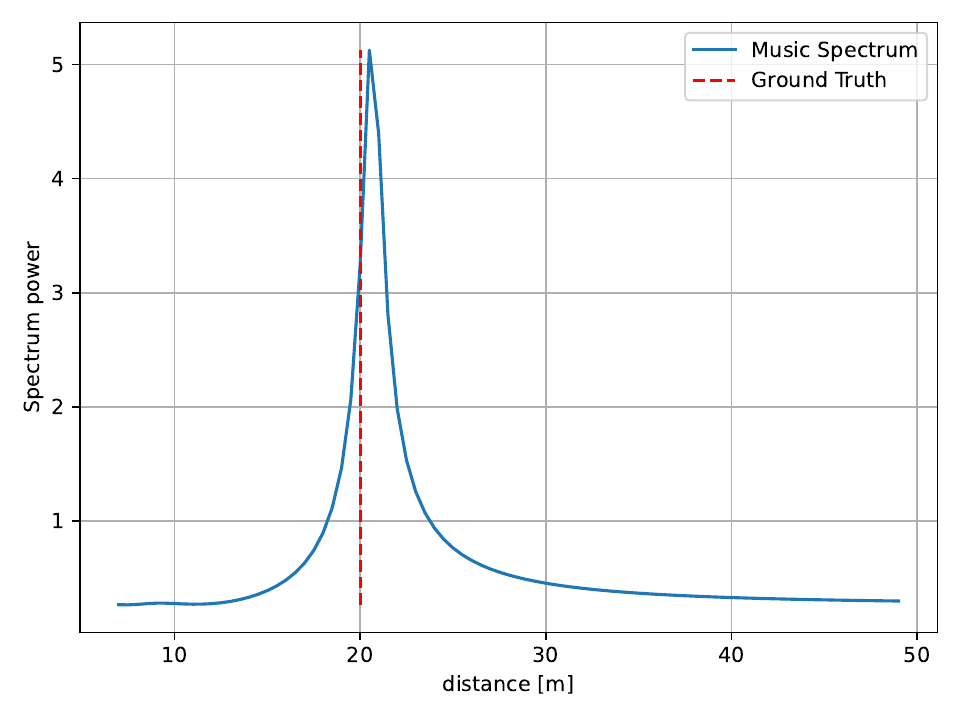}
            \caption{\ac{music} spec. at $\theta = 25^\circ$}
            \vspace{0.3cm}
        \end{subfigure}
    \end{minipage}
    
    \caption{\ac{music} spectrum for 2 non-coherent sources, located at ($-35^\circ$, $12m$), ($25^\circ$, $20m$). }
    \label{fig:spectrum non coherent}
\end{figure*}

\subsection{Interpretability}
\label{ssec:ExpInterp}
\begin{figure*}[t]
    \centering
    \begin{minipage}{0.70\textwidth} 
        \centering
        \begin{subfigure}[t]{0.49\textwidth}
            \centering
            \includegraphics[width=\linewidth]{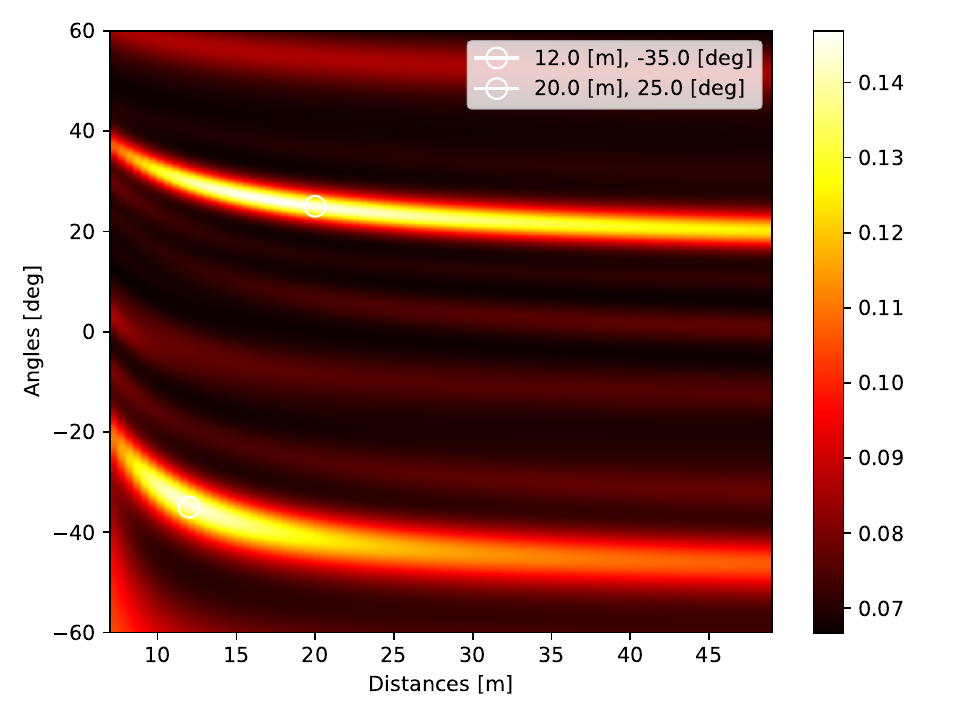}
            \caption{2D-\ac{music} spectrum}
        \end{subfigure}
        \hfill
        \begin{subfigure}[t]{0.49\textwidth}
            \centering
            \includegraphics[width=\linewidth]{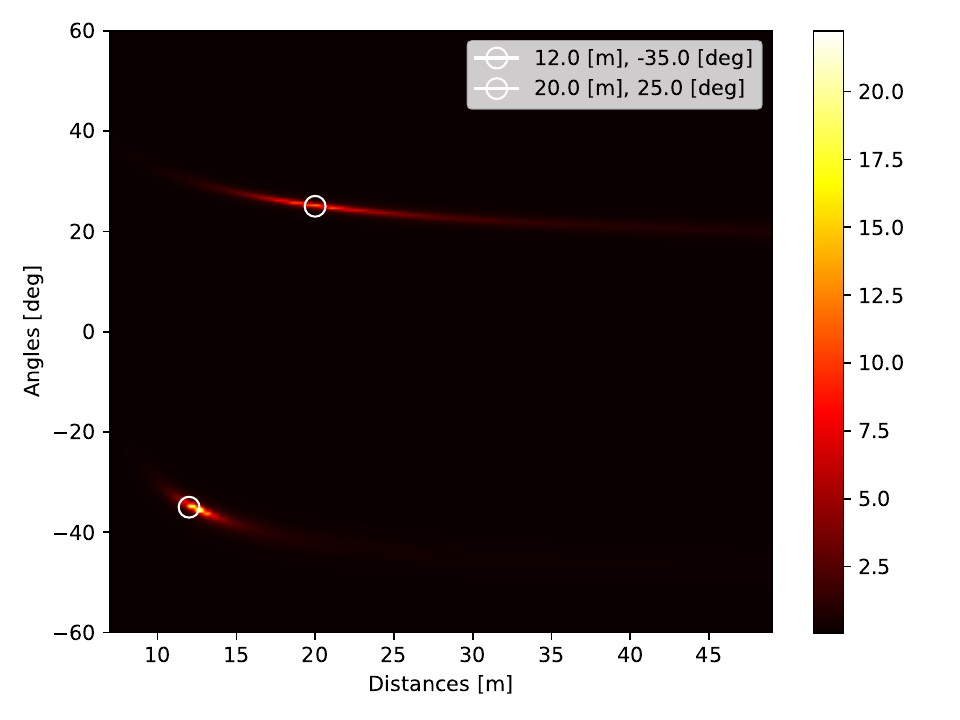}
            \caption{\name ~\ac{music} spectrum}
        \end{subfigure}
    \end{minipage}
    \hfill
    \begin{minipage}{0.25\textwidth} 
        \centering
        \begin{subfigure}[t]{\linewidth}
            \centering
            \includegraphics[width=\linewidth, height=3cm]{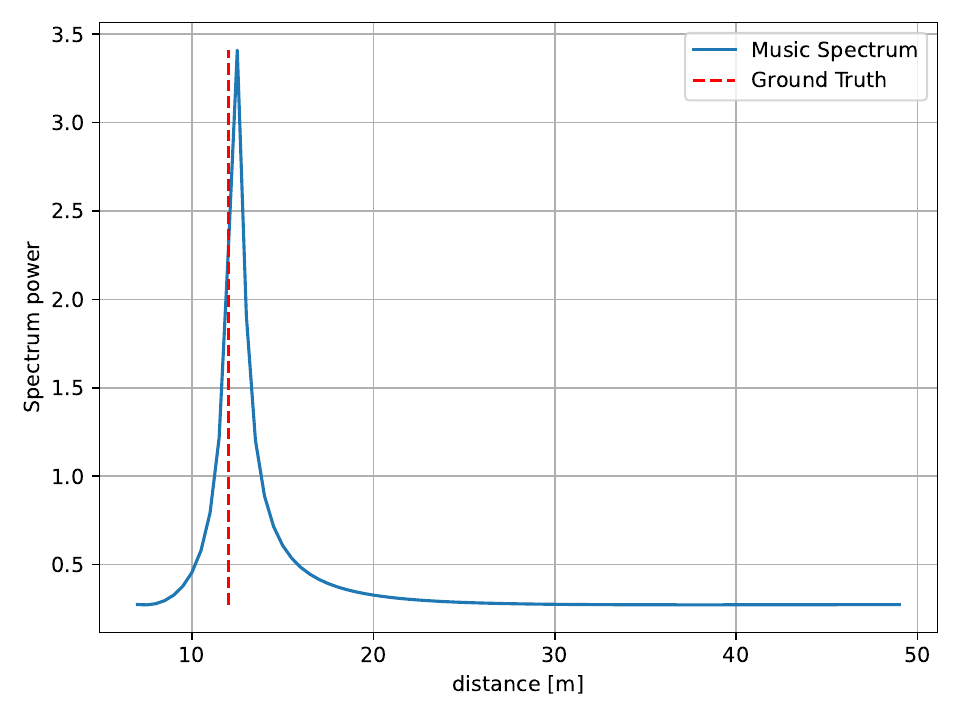}
            \caption{\ac{dcdmusic} spec at $\theta = -35^\circ$}
        \end{subfigure}
        
        \vspace{0.3cm} 
        
        \begin{subfigure}[t]{\linewidth}
            \centering
            \includegraphics[width=\linewidth, height=3cm]{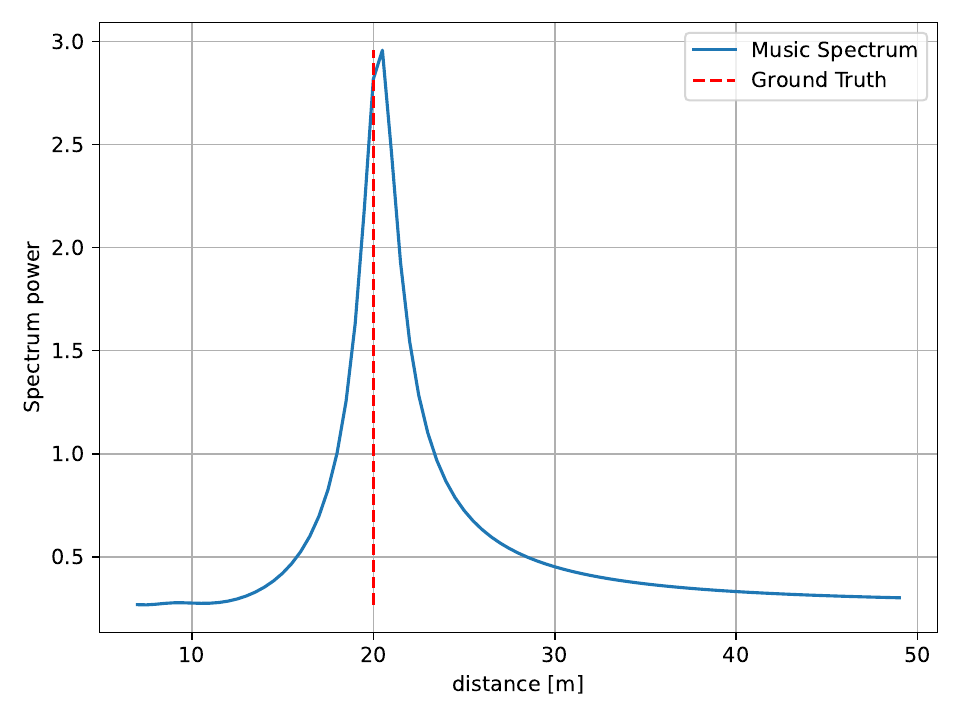}
            \caption{\ac{music} spec. at $\theta = 25^\circ$}
            \vspace{0.3cm}
        \end{subfigure}
    \end{minipage}
    
    \caption{spectra of 2D-\ac{music}, \name~and \ac{dcdmusic} for 2 coherent sources, located at ($-35^\circ$, $12m$), ($25^\circ$, $20m$).}
    \label{fig:spectrum coherent}
\end{figure*}

A major motivation for our usage of model-based deep learning is the desire to preserve the ability of subspace methods to provide an interpretable spectrum representation. To demonstrate this, we examine the spectrum of our methods, the 2D spectrum of \name~and 1D range  spectrum of \ac{dcdmusic}. We simulated $M = 2$ sources, with $10$ dB \ac{snr} and $T = 100$ snapshots for both coherent and non coherent scenarios. In Fig.~\ref{fig:spectrum non coherent} and Fig.~\ref{fig:spectrum coherent} we can see the spectrum for the non-coherent and coherent case respectively. Both \ac{dcdmusic} and \name~yield meaningful spectrum even when the sources are highly correlated, unlike the model-based 2D-\ac{music}. In addition, we  observe that the error grows with the sources approach to the Fraunhofer limit.

\begin{figure}
    \begin{subfigure}{0.24\textwidth}
        \includegraphics[width=\textwidth]{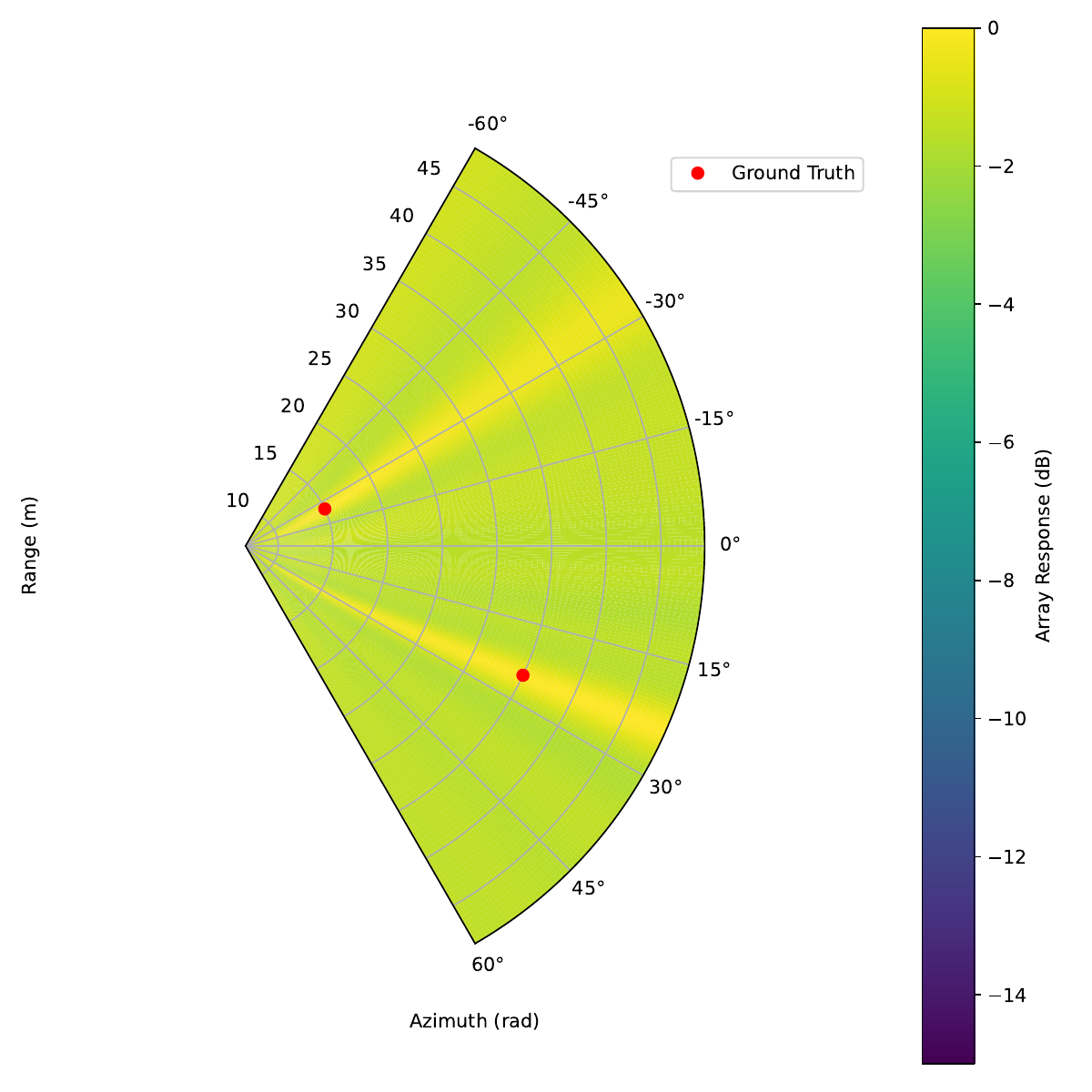}
        \caption{Sample covariance}
        \label{fig:2 coherent sources beampattern mvdr}
    \end{subfigure}%
    \hspace{0.05cm}
    \begin{subfigure}{0.24\textwidth}
        \includegraphics[width=\textwidth]{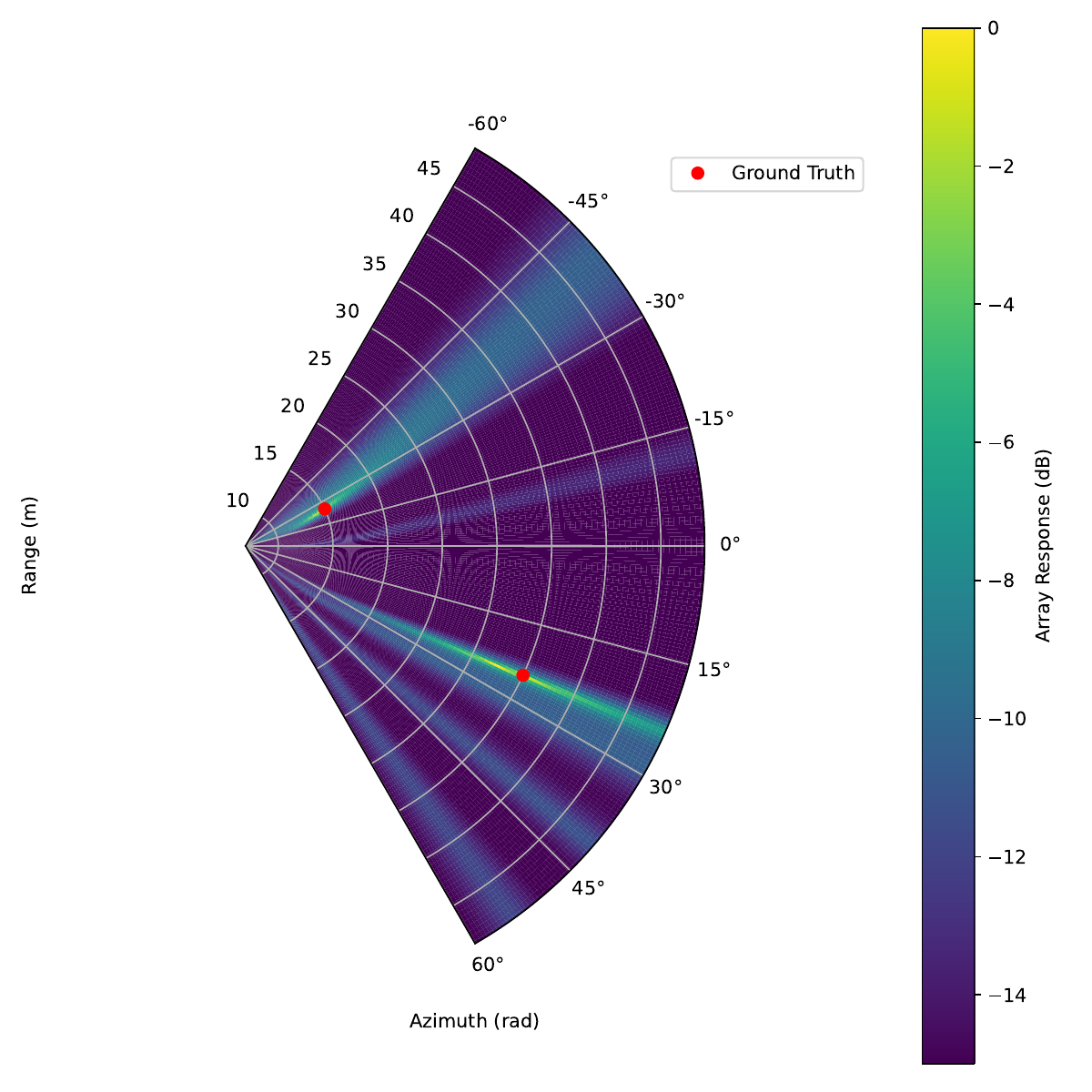}
        \caption{Augmented \name}
        \label{fig:2 coherent sources beampattern nfssn}
    \end{subfigure}
    \vspace{0.4cm}
    \caption{Beampattern for $M=2$ coherent sources.}
    \label{fig: 2 coherent sources beampattern}
\end{figure}

\begin{figure}
    \begin{subfigure}{0.24\textwidth}
        \includegraphics[width=\textwidth]{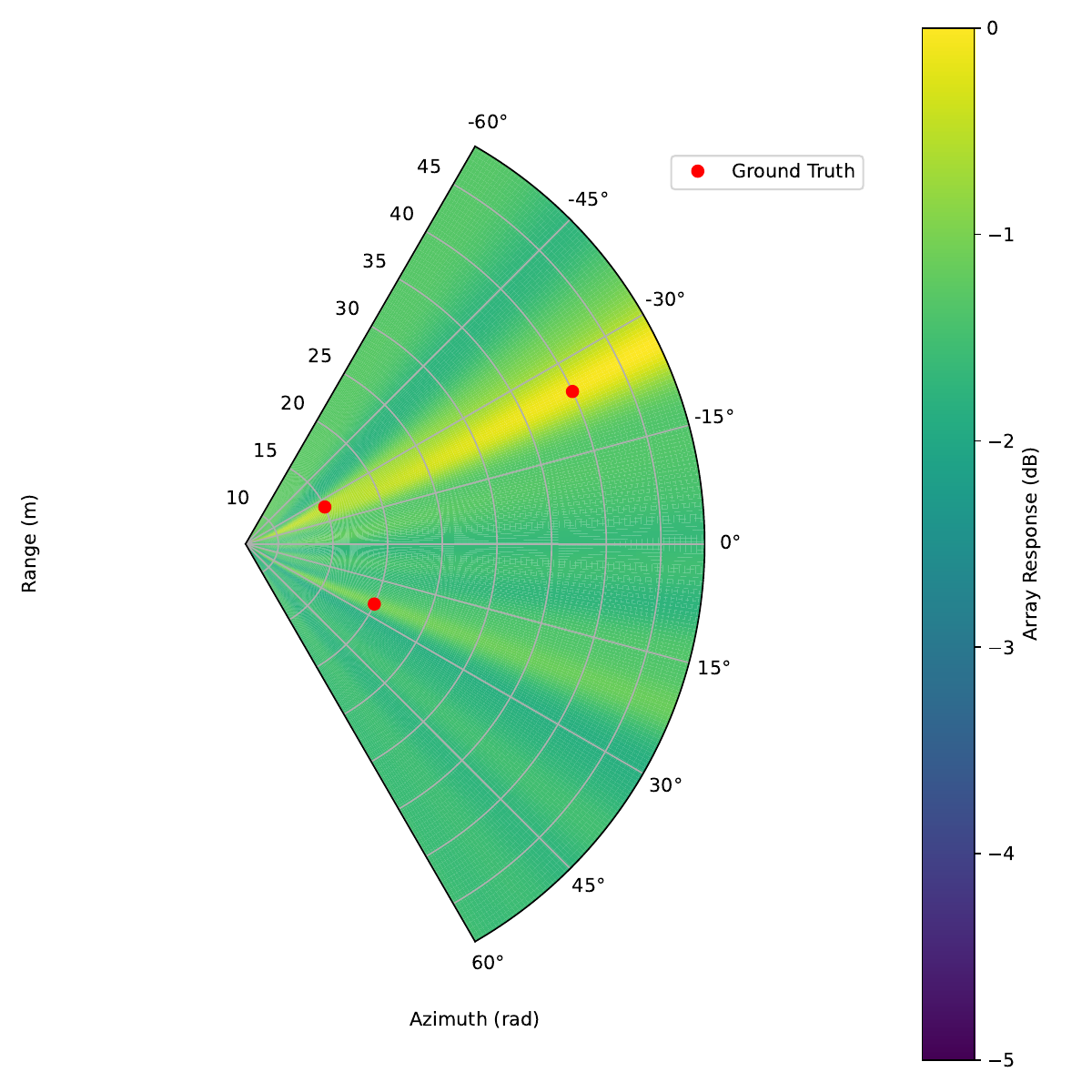}
        \caption{Sample covariance}
        \label{fig:3 coherent sources beampattern mvdr}
    \end{subfigure}%
    \hspace{0.05cm}
    \begin{subfigure}{0.24\textwidth}
        \includegraphics[width=\textwidth]{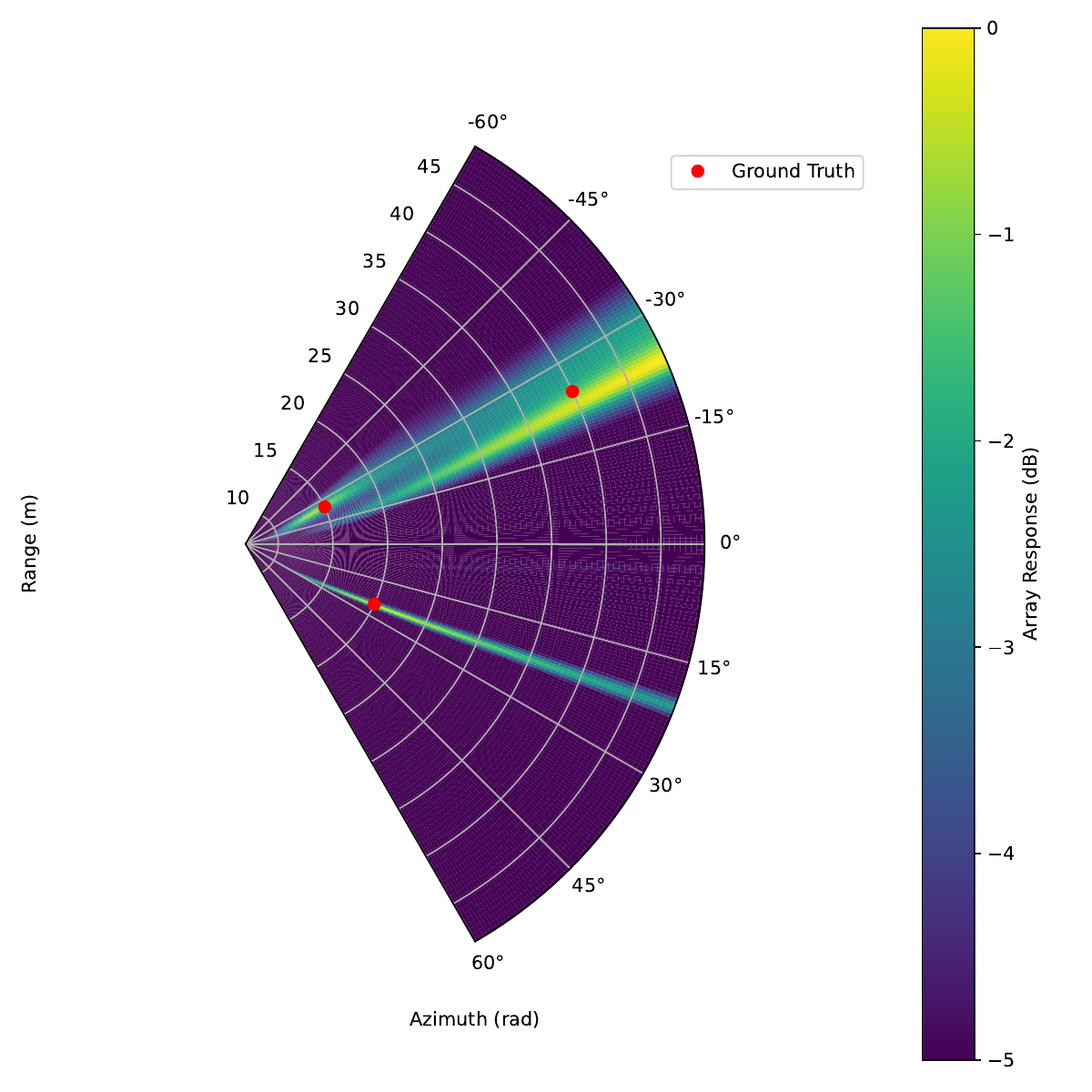}
        \caption{Augmented \name}
        \label{fig:3 coherent sources beampattern nfssn}
    \end{subfigure}
    \vspace{0.4cm}
    \caption{Beampattern for $M=3$ coherent sources.}
    \label{fig: 3 coherent sources beampattern}
\end{figure}

In addition to observing the \ac{music} spectrum, we also evaluate the usefulness of the surrogate near-field covariance produced by \name~for generating focused beams. To this end, we simulate two different scenarios. In the first scenario, we consider $M = 2$ coherent sources with \ac{snr} of $10$ dB, located at $(-25^\circ, 15 \text{ m})$ and $(25^\circ, 35 \text{ m})$. In the second scenario, we simulate $M = 3$ coherent sources with an \ac{snr} of $0$ dB, positioned at $(-25^\circ, 15 \text{ m})$, $(25^\circ, 20 \text{ m})$, and $(-25^\circ, 40 \text{ m})$. The latter scenario is significantly more challenging due to sources sharing the same angle. The results are presented in Fig.~\ref{fig: 2 coherent sources beampattern} and Fig.~\ref{fig: 3 coherent sources beampattern} for the two-source and three-source cases, respectively. These beampatterns highlight the advantages of our model-based deep learning approach, demonstrating that the surrogate covariance is a valuable tool for various tasks, including beamforming. When comparing to the classic approach, the surrogate covariance enables a clear distinction between coherent sources, even when they share the same angle. This is evidenced by the clear beam focusing around each source. In contrast, using sample covariance directly for beamforming  results in a more uniform energy distribution across different locations.

\subsection{Time Complexity}
We next complement the complexity analysis of Subsection~\ref{subsec:discussion}, and add an empirical time analysis in addition to assessing non-asymptotic complexity by counting \acp{mac}. The time analysis was performed by averaging the inference time of each model over $500$ trials from the setup of Fig.~\ref{fig: 2 sources results}, each passed with a batch size of 1. The \ac{mac} analysis was done using the THOP library\footnote{\url{https://pypi.org/project/thop/}}. The simulations were run on the same \textit{CPU}, without threading, over a \textit{Linux OS}, and ignoring any potential acceleration that could be achieved by using \textit{GPUs}. 
The results are shown in Table~\ref{tab:model_comparison}. We can observe that, although \name~has fewer \ac{mac} operations than \ac{dcdmusic}, its inference time is higher due to the expensive 2D grid search involved in its architecture. Furthermore, \ac{transmusic} has a slower inference time despite having fewer \ac{mac} operations, likely due to the use of the {Transformer} architecture, which is known for being more computationally intensive in terms of memory and operation complexity.

\begin{table}
    \centering
\adjustbox{max width =\columnwidth}{
    \centering
    \begin{tabular}{|c|c|c|}
        \hline
        \textbf{Model}        & \textbf{MACs (M)} & \textbf{Inference Time (ms)} \\ \hline
        \ac{dcdmusic}              & 29.383                               & \bgreen{5.691}                               \\ \hline
        \name         & \bgreen{14.692}                               & 15.160                              \\ \hline
        DeepCNN               & 212.774                              & 16.527                              \\ \hline
        \ac{transmusic}            & 19.843                               & 18.086                              \\ \hline
    \end{tabular}
    }
    
    \vspace{0.1cm}
    \caption{Latency and \ac{mac} comparison}
    \label{tab:model_comparison}
\end{table}

\section{Conclusions}
\label{sec:conclusions}
We proposed two \ac{ai}-aided subspace algorithms for near-field localization. The first algorithm, coined \name, is derived from Near-Field \ac{music}, while our second algorithm, \ac{dcdmusic}, augments cascaded \ac{esprit}-\ac{music} methods. We introduced dedicated training schemes for jointly assessing the usefulness of the \ac{music} spectrum, the accuracy in recovering the number of sources, and localization performance. Our experimental evaluations consistently demonstrate that our \ac{ai}-aided methods localize accurately in challenging settings, outperforming  model-based and data-driven benchmarks.


\bibliographystyle{IEEEtran}
\bibliography{IEEEabrv,mybib}


 





\end{document}